\documentclass{article}
\usepackage[dblblindworkshop,final]{neurips_2026}
\usepackage[utf8]{inputenc} % allow utf-8 input
\usepackage[T1]{fontenc}    % use 8-bit T1 fonts
\usepackage{hyperref}       % hyperlinks
\usepackage{url}            % simple URL typesetting
\usepackage{booktabs}       % professional-quality tables
\usepackage{amsfonts}       % blackboard math symbols
\usepackage{nicefrac}       % compact symbols for 1/2, etc.
\usepackage{microtype}      % microtypography
\usepackage{xcolor}         % colors

\usepackage{amsmath,amsfonts,bm}
\usepackage{tcolorbox}
\usepackage{makecell}

\newcommand{\globalM}[1]{\mathcal{M}_\theta^{#1}}
\newcommand{\genmodel}{\mathcal{M}_{\omega}}

\newcommand{\returns}{\rightarrow}

\newcommand{\traindata}{D}
\newcommand{\cleandata}{D^{c}}
\newcommand{\wmdata}{D^{w}}
\newcommand{\params}{\theta}
\newcommand{\globalparams}[1]{\params^{#1}}
\newcommand{\localparams}[2]{\params_{#1}^{#2}}

\newcommand{\secretkey}{s}
\newcommand{\true}{\textit{True}}
\newcommand{\false}{\textit{False}}
\newcommand{\signif}{\alpha}
\newcommand{\inputs}[1]{\textbf{x}_{#1}}
\newcommand{\nullhypo}{H_{0}}
\newcommand{\eval}[2]{\mathcal{E}(#1, #2)}

\newcommand{\agg}{\textsc{Agg}}
\newcommand{\fil}{\textsc{Fil}}

\newcommand{\watermark}[2]{\textsc{Watermark}^{#1}_{#2}}
\newcommand{\detect}[2]{\textsc{Detect}^{#1}_{#2}}

\newcommand{\adversary}{\mathcal{A}}
\newcommand{\vanilla}{\mathcal{T}}

\newcommand*\circled[1]{\tikz[baseline=(char.base)]{
            \node[shape=circle,draw,inner sep=1pt,minimum height=11pt] (char) {#1};}}

\newcommand{\vanillaFL}{\text{VanillaFL}\xspace}
\newcommand{\activeFL}{\text{ActiveFL}\xspace}

\newcommand{\wmworker}{k}      %
\newcommand{\wmratio}{\epsilon}             %
\newcommand{\allworker}{N}
\newcommand{\cleansigma}{\| \Sigma_C \|_2}
\newcommand{\updates}[2]{\Delta\params_{#1}^{#2}}
\newcommand{\setupdates}[1]{\{ \updates{i}{t} \}_{i=1}^{#1}}
\newcommand{\setclients}{C}
\newcommand{\wmupdates}{{W}_\Delta}
\newcommand{\cleanupdates}{{C}_\Delta}
\newcommand{\allupdates}{{U}_\Delta}

\newcommand{\temperature}{T}   %
\newcommand{\temperdiff}{\Delta{\temperature}}
\newcommand{\logitsbias}{\delta}            %

\newcommand{\prompt}{\pi}

\newcommand{\numlayer}{L}
\newcommand{\eachlayer}{\ell}
\newcommand{\wmworkerl}{\mathcal{W}_{\eachlayer}} 
\newcommand{\filterworkerl}{\mathcal{F}_{\eachlayer}} 
\newcommand{\wmfilterworkerl}{\wmworkerl \cap \filterworkerl}
\newcommand{\avgfilter}{\frac{1}{\numlayer} \sum_{\eachlayer=1}^{\numlayer}}
\newcommand{\evasionrate}{\mathrm{ER}}
\newcommand{\overfilteringrate}{\mathrm{OFR}}
\newcommand{\clientopt}{\textsc{ClientOpt}}
\newcommand{\serveropt}{\textsc{ServerOpt}}
\newcommand{\outlierremoval}{\textsc{OutlierRemovalSubroutine}}
\newcommand{\Kirchenbauer}{\textnormal{KGW+}}
\newcommand{\Kuditipudi}{\textnormal{KTH+}}
\newcommand{\ITS}{\textnormal{KTH-ITS}}
\newcommand{\EXP}{\textnormal{KTH-EXP}}
\newcommand{\pval}{\textnormal{$p$-value}}
\newcommand{\pvals}{\textnormal{$p$-values}}

\newcommand{\cfour}{\textnormal{C4}}
\newcommand{\alpaca}{\textnormal{Alpaca}}

\newcommand{\randeigen}{\textnormal{RandEigen}}
\newcommand{\FedOpt}{\textnormal{FedOpt}}

\newtcolorbox{embox}{
    colback=white,
    coltext=black,
    boxrule=1pt,
    width=\linewidth,
    before skip=0.5\baselineskip,
    after skip=0.5\baselineskip,
}

\newtheorem{definition}{Definition}

\newcommand{\myparatight}[1]{\smallskip\noindent{\bf {#1}.}~}

\usepackage{enumitem}
\usepackage{amsmath}
\usepackage{subcaption}
\usepackage{algorithm}
\usepackage{xspace}
\usepackage{graphicx}
\usepackage{wrapfig}
\usepackage{tikz}
\usepackage[noend]{algpseudocode}
\hypersetup{hidelinks}
\usepackage{siunitx}
\title{Watermark Robustness and Radioactivity May Be at Odds in Federated Learning}
\workshoptitle{3rd Workshop on Attributing Model Behavior at Scale: Data Attribution and Provenance}

\author{%
  Leixu Huang \qquad Zedian Shao \qquad Teodora Baluta \\
  Georgia Institute of Technology \\
  \texttt{\{lhuang340,zedian.shao,teobaluta\}@gatech.edu}
}

\begin{document}
\maketitle
\begin{abstract}
 Federated learning (FL) enables fine-tuning large language models (LLMs) across distributed data sources. As these sources increasingly include LLM-generated text, provenance tracking becomes essential for accountability and transparency. We adapt LLM watermarking for FL data provenance, where a subset of clients computes local updates on watermarked data, and the server averages all updates into the global LLM. We find that several evaluated LLM watermarks remain {\em radioactive} in this setup: the watermark signal stays detectable after fine-tuning, with a $\pval$ reaching $10^{-24}$ even when only $6.6\%$ of data is watermarked. However, the server can act as an {\em active adversary}, aiming to preserve model utility while evading provenance tracking. We observe that updates from watermarked synthetic data appear as outliers relative to clean updates. Our adversary thus applies strong robust aggregation to filter out these outliers along with the watermark signal. All evaluated radioactive watermarks are {\em not robust} against such an adversarial server. Our work suggests fundamental trade-offs between radioactivity, robustness, and utility.
\end{abstract}

\section{Introduction}
\label{sec:intro}
Large language models (LLMs) are increasingly used to generate synthetic datasets for fine-tuning, motivated by the high cost of human annotations and expert knowledge~\citep{alpaca-stanford,wang2023self,li2023synthetic}. Despite improving model generalization and mitigating misalignment in sensitive domains including healthcare~\citep{nik2023generation,nikolentzos2023synthetic} and finance~\citep{harsha2025synthetic}, synthetic data are often restricted from direct sharing due to privacy concerns and regulations like GDPR~\citep{voigt2017eu} and HIPAA~\citep{act1996health}.

Federated learning (FL) enables fine-tuning across distributed sources by training locally and aggregating updates into a global model~\citep{kairouz2021advances,rieke2020future,tian2022fedbert,caldas2018leaf}. While this mitigates data exposure, it does not address data provenance, i.e., attributing data contributions to their providers. Recent AI regulations recognize provenance as key to accountability and transparency~\citep{ai_act_2024}, yet it remains unexplored in FL.

Watermarking offers a potential mechanism for provenance by embedding a secret statistical signal in LLM-generated data. Prior work shows that LLM watermarks exhibit {\em radioactivity} in centralized training, i.e. the signal remains detectable in a model fine-tuned on watermarked text~\citep{sander2024watermarking, gu2024learnability}. In FL, however, local-training noise and client drift may dilute the watermark signal~\citep{shi2022optimization}. It is unclear whether existing watermarks remain radioactive in FL.

Moreover, the FL server may act as an {\em active adversary}, deliberately attempting to evade provenance tracking. This is a new threat model for watermark {\em robustness}. Unlike in non-FL setup, the FL server lacks access to clean data and does not control the fraction of watermarking clients. It must also preserve the global LLM utility, so it cannot arbitrarily discard updates to remove the watermark. This raises the important question: Does there exist an effective attack to remove watermarks in FL without sacrificing model utility?

\begin{wrapfigure}{r}{0.4\textwidth}
    \centering
    \vspace{-10pt}
    \includegraphics[width=0.4\textwidth]{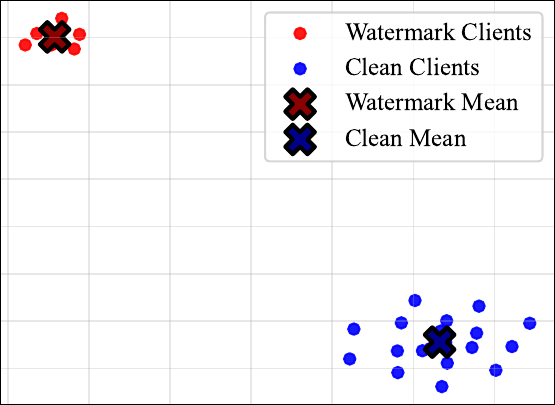}
    \caption{t-SNE visualization of model updates shows the clear differences in updates from clean clients (blue) and watermark clients (red). 
    }
    \vspace{-35pt}
    \label{fig:gradient_distributions}
\end{wrapfigure}

In this paper, we show that existing watermarks are radioactive in FL but removable by an active server. The key observation is that updates from clean and watermarking clients are clearly separable, as shown in Figure~\ref{fig:gradient_distributions} using t-SNE~\citep{maaten2008visualizing}. Watermark removal thus reduces to a robust aggregation problem~\citep{diakonikolas2017being,choudhary2024attacking,lee2025practical}: an adversary can simply filter out outliers, i.e., updates falling outside the variance of the clean update distribution. We show that all evaluated watermarks are not robust against such attack.

To the best of our knowledge, this is the first work to introduce federated data provenance (Section~\ref{sec:fed-prov-problem}).
In summary, our contributions are the following: 

\begin{list}{\labelitemi}{\leftmargin=2em \itemindent=-0.3em \itemsep=.2em}
    \item We adapt existing watermarks to benign federated LLM fine-tuning~\citep{reddi2020adaptive}, and empirically show that {\em radioactivity remains in FL}, with $\pvals$ ranging from $10^{-3}$ to $10^{-24}$.
    \item We formulate an active adversary threat model and show that a strong robust aggregator filters out watermarked updates, causing radioactivity detection to fail in {\em all evaluated setups}.
    \item We provide an extensive evaluation showing that none of the current LLM watermarks achieve {\em radioactivity, robustness} and {\em utility} simultaneously across our evaluated setups.
\end{list}

\section{Preliminaries}
\label{sec:background}

\subsection{Federated Setup}
\label{sec:background-fl}
In the federated fine-tuning setup, $\allworker$ clients $\setclients = \{c_1, \dots, c_N\}$ collaboratively train a global LLM $\globalM{}$ under the coordination of a central server. $\globalM{}$ has model parameters $ \params \in \mathbb{R}^d$, where $d$ is the number of parameters.
Each client $c_i$ maintains a private dataset $\traindata_i$. At each communication round $t$, the server broadcasts $\globalparams{t}$ to all clients. Each client fine-tunes on $\traindata_i$ and returns a local update $\updates{i}{t}$. The server aggregates clients updates via an aggregation function $\agg: {\mathbb{R}}^{\allworker \times d} \rightarrow {\mathbb{R}}^d$ to update the global model $\globalparams{t+1} = \globalparams{t} + \agg (\{ \updates{i}{t} \}_{i=1}^N)$. Let $\globalM{}$ denote the final model after training.

\subsection{LLM Watermarking}
\label{sec:background-wm}
Let $\genmodel$ be an LLM that generates text autoregressively from a prompt $\prompt$. 
To watermark the outputs of $\genmodel$, a watermarking function $\watermark{\genmodel}{\secretkey}(\prompt) \rightarrow \inputs{}^w$ uses a secret key $\secretkey$ to perturb decoding, embedding a detectable signal into the output $\inputs{}^w$~\citep{zhao2025sok}. 
$\Kirchenbauer$~\citep{kirchenbauer2023watermark} is a hashing-based watermark that biases the sampling towards a pseudo-random subset of vocabulary. $\Kuditipudi$~\citep{kuditipudi2023robust} instead pre-defines a random number sequence with $\secretkey$ and preserves the output distribution of $\genmodel$. A detection function $\detect{}{\secretkey}(\inputs{})$ performs a statistical test on $\inputs{}$ using $\secretkey$ to determine whether $\inputs{}$ was watermarked. 

\myparatight{Radioactivity} Let $\genmodel$ be the model that generates a watermarked dataset $\wmdata$. To evaluate whether another model $\globalM{}$ has fine-tuned on $\wmdata$, a modified detection function $\detect{\globalM{}}{\secretkey}(\wmdata)$ is used. This variant examines $\globalM{}$'s predictions on $\wmdata$ rather than $\wmdata$ itself~\citep{sander2024watermarking}. The detection accumulates a score over $\globalM{}$'s predictions and performs a statistical test $T$ against a null distribution to obtain a $p$-value. The $p$-value indicates how likely the score is under the null hypothesis that $\globalM{}$ was not fine-tuned on $\wmdata$. Comparing $p$-value to a predefined significance level $\signif$ yields a binary decision: $\true$ or $\false$. See Appendix~\ref{appendix:llm-wm-details} for more details on LLM watermarking. 
\section{Federated Data Provenance}
\label{sec:fed-prov-problem}

We study data provenance in FL via watermarking. An $\wmratio$ fraction of clients aim to prove that their data contribute to the global model $\globalM{}$. Each such client watermarks its local data with the generation method $\watermark{\genmodel}{\secretkey}$, producing $\wmdata_i$. At round $t$, all clients send local updates $\allupdates=\setupdates{N}$ to the server for aggregation. We denote updates from watermarked data as $\wmupdates$ and clean updates as $\cleanupdates$. After training, watermarking clients verify their contribution to $\globalM{}$ using $\detect{\globalM{}}{s}$.

We consider two server settings. In the vanilla setup (\textbf{\vanillaFL}), the server simply averages updates from all clients, $\vanilla(\allupdates,\globalparams{t}) =\globalparams{t} + \frac{1}{N} \sum_{i=1}^N \updates{i}{t}$~\citep{reddi2020adaptive}. Here, the watermark's only obstacle is surviving dilution from noise and client drift~\citep{shi2022optimization}, making \vanillaFL the radioactivity baseline. In contrast, an adversarial server lacks incentive to preserve provenance signals. We refer to this as {\em active adversary} setup (\textbf{\activeFL}), where the server aims to evade watermark detection while maintaining the LLM utility.

\myparatight{Threat Model} 
We assume no privacy attacks, where neither the server nor other clients try to infer a client's local data. In \vanillaFL, the server follows the FL protocol, averaging updates. In \activeFL, the server may use different aggregation methods but must otherwise follow the protocol. Following~\citet{lee2025practical}, we assume that the server knows the watermark ratio $\wmratio$. All clients are honest, keep their data private, and follow the FL protocol. We assume watermarking clients have a shared key $\secretkey$ and generative model $\genmodel$ unknown to the server. More details are in Appendix~\ref{appendix:ps}.

\subsection{Problem Statement}
\label{sec:problem-statement}
We consider federated data provenance under the threat model of an active adversary $\adversary$. $\adversary$ takes as input the clients updates $\allupdates$ ($\wmratio < 0.5$) and the current global model parameters $\globalparams{t}$, returning the updated parameters $\globalparams{t+1}$. The adversary in \activeFL aims to $\circled{1}$ obtain $\globalM{t+1}$ matching the utility $\vanilla$ would achieve using only clean updates and $\circled{2}$ reduce watermark detectability, i.e., increase the $p$-value obtained from testing $\globalM{t+1}$s prediction on $\wmdata_i$. 

The formal definition of watermark robustness under FL is in Appendix~\ref{appendix:ps}. Intuitively, it is \emph{counterfactual}, similar to its counterpart for watermarked LLM generations~\citep{zhao2025sok}: it asks whether $\wmdata_i$, which would be radioactive had the model been updated with $\vanilla$, remains radioactive when $\adversary$ updates the model instead. The clients and server have conflicting goals regarding watermark detection, yet all parties aim to preserve model utility. Our paper addresses whether such an active adversarial server exists.

\section{Approach}
\label{sec:approach}

\begin{wrapfigure}{r}{0.45\textwidth}
    \centering
    \vspace{-10pt}
        \includegraphics[trim=38mm 45mm 100mm 8mm, clip, width=0.45\textwidth]{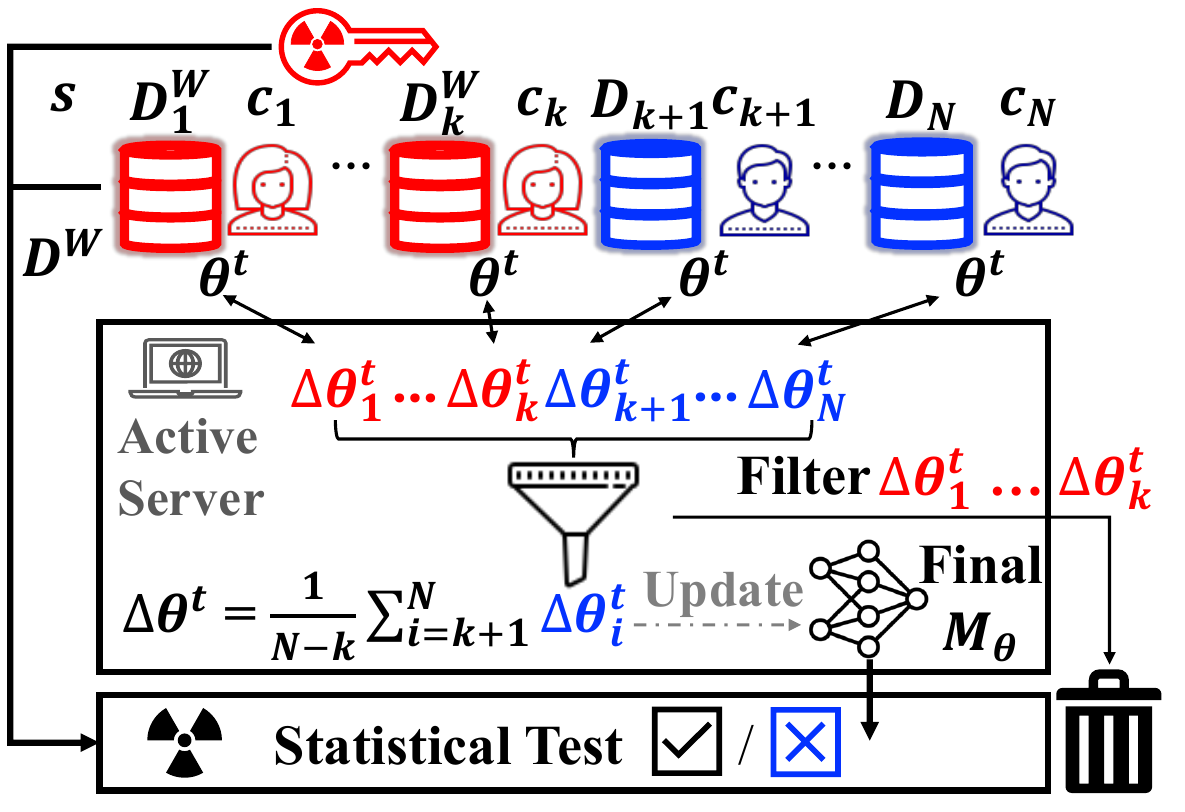}
        \caption{Overview of federated data provenance in \activeFL for LLMs.}
        \label{fig:system_overview}
    \vspace{-10pt}
\end{wrapfigure}

Figure~\ref{fig:system_overview} presents an overview of \activeFL, where we propose an active server $\adversary$ that removes the watermarked updates by filtering the $\wmupdates$ from $\allupdates$ at each round $t$. The challenge is in distinguishing watermarked updates from clean ones as the server must preserve enough $\updates{i}{t} \in \cleanupdates$ for effective learning. Our insight is that LLM watermarks like $\Kirchenbauer$ are radioactive but low-distortion, i.e., there is a small statistical distance between the watermarked and clean data distributions~\citep{zhao2025sok}. If distortions from watermarked LLM-generated text propagate to updates computed on such texts, then there is a {\em measurable shift} in the watermarked vs. clean updates distributions. Specifically, we observe that watermarked updates $\updates{i}{t} \in \wmupdates$ become outliers in the distribution of clean updates $\cleanupdates$. By removing outliers from client updates before averaging, the server can thus satisfy goals~$\circled{1}$ and~$\circled{2}$ simultaneously. Therefore, we employ strong Byzantine-robust aggregators that are designed to filter out outlier updates in our $\activeFL$ setup.

Byzantine robust aggregation estimates the true mean of model updates from clean clients while mitigating the influence of corrupted ones~\citep{Huber1964RobustEO}. Such aggregator $\fil$ takes a set $\allupdates$ of $N$ vectors in $\mathbb{R}^d$ (local updates from $N$ clients), where an $\wmratio$-fraction are arbitrarily corrupted (watermarked). $\fil$ aims to minimize the distance between its output and the true mean of the uncorrupted (clean) subset $\cleanupdates \subseteq \allupdates$, which we denote as \textit{bias}. \emph{Strong} aggregators bound this bias by $O(1)$, independent of vector dimension $d$~\citep{diakonikolas2017being}. This is essential given the very large dimensionality of LLM updates. Since such aggregators bound the bias by iteratively removing corrupted (watermarked) updates, they can eventually cause watermark detection to fail. Full guarantees and algorithm details for robust aggregation can be found in Appendix~\ref{appendix:Byzantine-robust-agg}. 

Based on these insights, we develop a framework for integrating data provenance into FL fine-tuning for both \vanillaFL and \activeFL. Our framework builds upon \FedOpt~\citep{reddi2020adaptive}, with two key modifications. First, we broadcast the global model to \emph{all} clients each round, rather than to a subset, to accelerate convergence. Second, the aggregation step $\agg$ is either simple averaging ($\vanillaFL$) or a strong Byzantine-robust aggregator ($\activeFL$), depending on the server type. After aggregating with $\agg$, the server updates the global model by applying $\serveropt$ on the aggregated results. Fine-tuning uses early stopping, i.e., training terminates when the evaluation loss ceases to decrease. Upon completion, clients perform watermark detection on the final global model $\globalM{t}$. The complete algorithm for the framework (Algorithm~\ref{alg:wm_in_fl}) is provided in Appendix~\ref{appendix:imple-setup-details}.
\section{Evaluation}
\label{sec:eval}

\begin{table}[t]
\centering
\begin{minipage}[t]{0.48\textwidth}
    \centering
    \vspace{0pt} 
    \setlength{\tabcolsep}{2pt}
    \caption{LLM watermarks radioactivity under FL fine-tuning with $\wmratio = 6.6\%$. Pre-fine-tuning $\pval \sim 0.5$, consistent with $\nullhypo$. $\Kirchenbauer$ shows strong and model-size-dependent radioactivity. $\ITS$ shows no radioactivity.}
\centering
\footnotesize
\renewcommand{\arraystretch}{0.92}
\begin{tabular}{ccccc}
\toprule
\textbf{Data} & \textbf{WM} & \textbf{Model} & \multicolumn{2}{c}{\textbf{$\bm{p}$-value}}\\
\cmidrule(lr){4-5}
& & & \textbf{Before FT} & \textbf{After FT} \\
\midrule
$\cfour$    &   $\Kirchenbauer$ & 70M  & 0.584 & $1.69 \times 10^{-1}$ \\
            &                   & 160M & 0.397 & $1.27 \times 10^{-3}$ \\
            &                   & 410M & 0.877 & $2.41 \times 10^{-8}$ \\
\cmidrule(lr){2-5}
            &   $\ITS$   & 70M  & 0.485 & 0.500      \\
            &                   & 160M & 0.500 & 0.500      \\
            &                   & 410M & 0.480 & 0.480      \\
\midrule
$\alpaca$   &   $\Kirchenbauer$ & 70M  & 0.204 & $1.30\times 10^{-2}$  \\
            &                   & 160M & 0.309 & $1.59\times 10^{-11}$ \\
            &                   & 410M & 0.302 & $4.96\times 10^{-24}$ \\ 
\cmidrule(lr){2-5}
            &   $\ITS$   & 70M  & 0.490 & 0.480      \\
            &                   & 160M & 0.480 & 0.480       \\
            &                   & 410M & 0.480 & 0.480    \\
\bottomrule
\end{tabular}

    \label{tab:baseline_radioactive}
\end{minipage}
\hfill
\begin{minipage}[t]{0.48\textwidth}
    \centering
    \vspace{0pt} 
    \includegraphics[width=0.94\textwidth]
    {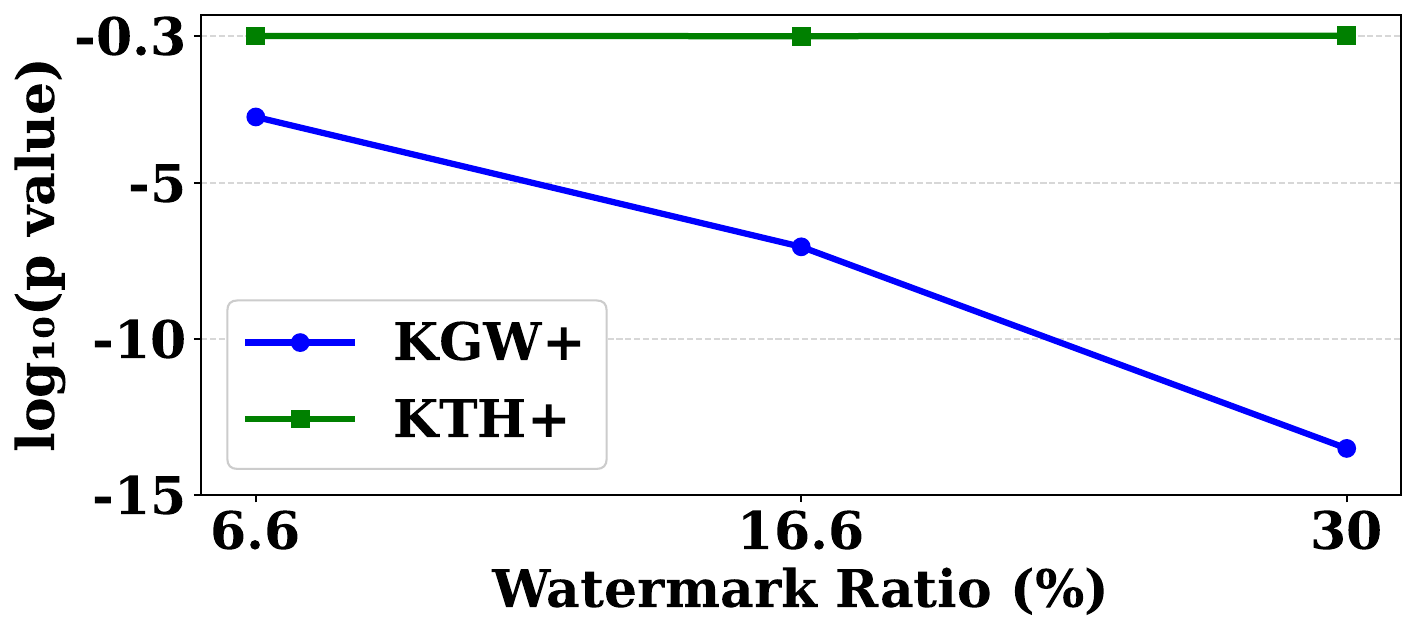}
    \captionof{figure}{$\Kirchenbauer$ radioactivity improves with larger $\wmratio$, while $\ITS$ remains not radioactive.}
    \label{tab:eps_radioactive}
    
    \vspace{2pt}
    
    \includegraphics[width=0.94\textwidth]{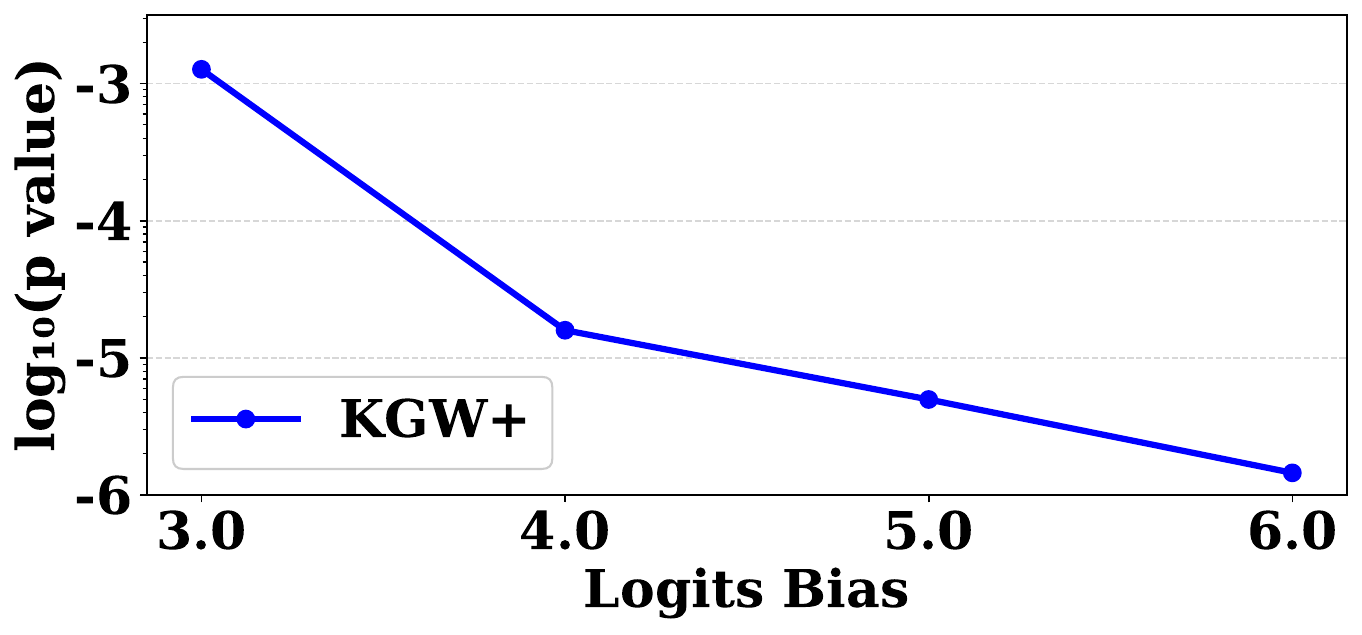}
    \captionof{figure}{$\Kirchenbauer$ radioactivity improves with larger $\logitsbias$.}
    \label{tab:delta_radioactive}
\end{minipage}
\vspace{-14pt}
\end{table}

In this section, we ask the following research questions:
\begin{enumerate}[noitemsep,label=\textbf{(RQ{\arabic*})}]
    \item Are LLM watermarks radioactive in the federated learning settings? 
    \item Are LLM watermarks robust against strong Byzantine robust aggregation? 
    \item How do watermark hyper-parameters affect the radioactivity/robustness trade-off?
    \item Can watermark radioactivity and robustness be achieved simultaneously? If so, does this introduce a broader trade-off with utility?
\end{enumerate}

\subsection{Setup}
We briefly outline the FL setup, watermarking schemes, and metrics. Full details in Appendix~\ref{subsec:imple-setup-details}.

\myparatight{Experimental Setup} We fine-tune Pythia~\citep{biderman2023pythia} (70M/160M/410M) on C4~\citep{raffel2020exploring} and Alpaca~\citep{alpaca-stanford}. Ablations use Pythia-160M and C4. We simulate 30 clients, $\wmworker \in \{2, 5, 9\}$ of which apply watermark ($\wmratio = \frac{\wmworker}{30} \in \{6.6\%, 16.6\%, 30.0\%\}$). $\activeFL$ uses $\randeigen$~\citep{lee2025practical} to aggregate, with early stopping after three rounds of worsening validation loss.
We evaluate $\Kirchenbauer$~\citep{kirchenbauer2023watermark} and $\ITS$~\citep{kuditipudi2023robust}, generated by Pythia-2.8b. $\Kirchenbauer$ uses softmax temperature $\temperature = 0.8$ and logit bias $\logitsbias = 3.0$.

\myparatight{Evaluation Metrics}
We measure global model utility with entropy loss and watermark robustness with \textbf{Evasion Rate ($\bm{\evasionrate}$)}, the fraction of watermarking clients surviving aggregation, and \textbf{Overfiltering Rate ($\bm{\overfilteringrate}$)}, the fraction of filtered clients that are \emph{not} watermarked. 

\subsection{LLM Watermark Radioactivity in FL}
\label{sec:eval-radioactivity}

We first evaluate LLM watermark radioactivity by fine-tuning $\globalM{}$ under $\vanillaFL$. We report $\pval$ accumulated across all watermarked datasets for simplicity (see Appendix~\ref{appendix sub:individual-detection} for details).

\myparatight{Watermark Radioactivity} $\Kirchenbauer$ exhibits strong radioactivity. Table~\ref{tab:baseline_radioactive} shows $\pval$ as low as $10^{-8}$ on $\cfour$ and $10^{-24}$ on $\alpaca$ with only 6.6\% watermarked samples. We expect this extends to hashing-based methods using the same cumulative detector~\citep{fu2024gumbelsoft, lee2024wrote, aaronson2023watermarking}. In contrast, $\ITS$ is not radioactive in our FL setup ($\pval \approx 0.5$), consistent with its weaker detectability reported in prior work~\citep{kuditipudi2023robust}. 

\myparatight{Influence of $\globalM{}$ Size and $\wmratio$} $\Kirchenbauer$ radioactivity improves with global model size. Table~\ref{tab:baseline_radioactive} shows that on $\cfour$, increasing model size from 70M to 410M parameters shifts detection from ineffective ($\pval = 0.169 > 0.01$) to effective ($\pval = 10^{-8} < 0.01$). $\Kirchenbauer$ radioactivity also improves with larger $\wmratio$. Figure~\ref{tab:eps_radioactive} shows that for Pythia-160M, post-finetuning $\pval$ drops from $10^{-3}$ to $10^{-14}$ as $\wmratio$ increases from 6.6\% to 30.0\%. In contrast, $\ITS$ remains not radioactive ($\pval \approx 0.5$) regardless of model size or $\wmratio$, further confirming its weak detectability.

\begin{embox}
\textbf{(RQ1)} Watermark radioactivity observed in the centralized setting also holds in FL. $\Kirchenbauer$ radioactivity improves with larger $\wmratio$ and $\globalM{}$ size.
\end{embox}

\subsection{LLM Watermark Robustness in FL}
\label{sec:eval-robustness}

We evaluate watermark robustness by comparing post-training $\pval$ under $\activeFL$ and $\vanillaFL$. We only consider configurations radioactive under $\vanillaFL$ ($\Kirchenbauer$ on Pythia-160M/410M). 

\myparatight{KGW+ Robustness} $\Kirchenbauer$ is not robust to $\randeigen$. Table~\ref{tab:baseline_robust} shows that $\vanillaFL$ yields low post-finetuning $\pval \leq 1.27 \times 10^{-3}$, whereas fine-tuning under $\randeigen$ yields $\pval \approx 0.5$. This indicates that $\randeigen$ effectively filters watermarked updates during aggregation, preventing reliable detection in the final global model. We expect this extends to other hashing-based watermarks~\citep{fu2024gumbelsoft, lee2024wrote, aaronson2023watermarking}, whose repetitive hashing windows should similarly result in consistent biases.

\myparatight{Influence of $\wmratio$} $\Kirchenbauer$ robustness improves with larger $\wmratio$. Table~\ref{tab:eps_robust} shows that $\evasionrate$ rises from 3.5\% to 9.9\%, as $\wmratio$ increases from 6.6\% to 30.0\%. Higher $\evasionrate$ indicates that more watermarking clients survive aggregation, thereby enhancing watermark robustness. However, $\Kirchenbauer$ remains undetectable ($\pval \approx 0.5$) across all evaluated $\wmratio$ under $\activeFL$. Thus, neither evaluated watermark is effective in FL: $\ITS$ is not radioactive, and $\Kirchenbauer$ is not robust to server's active filtering.

\begin{embox}
\textbf{(RQ2)} $\Kirchenbauer$ watermark is not robust against strong Byzantine robust aggregation. $\Kirchenbauer$ robustness improves with larger $\wmratio$ but remains undetectable even when $\wmratio$ reaches $30.0\%$.
\end{embox}

\begin{table}[t]
\centering
\subcaptionbox*{}
{\begin{minipage}[t]{0.42\textwidth}
    \centering
    \setlength{\tabcolsep}{5pt}
    \caption{$\Kirchenbauer$ robustness under $\activeFL$ with $\wmratio = 6.6\%$. $\Kirchenbauer$ is not robust against $\randeigen$.}
\label{tab:baseline_robust}
\centering
\footnotesize
\renewcommand{\arraystretch}{0.92}
\begin{tabular}{cccc}
\toprule
\textbf{Data} & \textbf{Model} & \multicolumn{2}{c}{\textbf{$\bm{p}$-value}} \\
\cmidrule(lr){3-4}
& & \textbf{Vanilla} & \textbf{Active}\\
\midrule
$\cfour$    & 160M & $1.27 \times 10^{-3}$ & 0.550 \\
            & 410M & $2.41 \times 10^{-8}$ & 0.613 \\
\midrule
$\alpaca$   & 160M & $1.59\times 10^{-11}$ & 0.231 \\
            & 410M & $4.96\times 10^{-24}$ & 0.282 \\
\bottomrule
\end{tabular}

\end{minipage}}
\hfill
\subcaptionbox*{}
{\begin{minipage}[t]{0.54\textwidth} 
    \centering
    \setlength{\tabcolsep}{5pt}
    \caption{
$\Kirchenbauer$ robustness against $\randeigen$ with varying watermark proportion ($\wmratio$).
Robustness improves (higher $\evasionrate$) with a larger $\wmratio$ but remain undetectable across all evaluated settings.}
\label{tab:eps_robust}
\centering
\footnotesize
\renewcommand{\arraystretch}{0.92}
\begin{tabular}{ccccc}
\toprule
\textbf{$\bm{\wmratio}$} & \multicolumn{2}{c}{\textbf{$\bm{p}$-value}} & \textbf{ER} & \textbf{OFR} \\
\cmidrule(lr){2-3}
& \textbf{Vanilla} & \textbf{Active} & (\%) & (\%) \\
\midrule
6.6\%   & $1.27 \times 10^{-3}$ & 0.550 & 3.5 & 48.2 \\
16.6\%  & $8.83 \times 10^{-8}$ & 0.517 & 8.4 & 53.6 \\
30.0\%  & $3.10 \times 10^{-14}$ & 0.733 & 9.9 & 50.1 \\
\bottomrule
\end{tabular}

\end{minipage}}
\vspace{-34pt}
\end{table}

\subsection{Watermark Hyperparameters Analysis}
\label{subsec:hyperparameters}

We test how three hyperparameters affect the radioactivity/robustness trade-off: the generative model ($\genmodel$), logits bias ($\logitsbias$), and the softmax temperature difference between clean and watermarked data ($\temperdiff$). Full setup details and tables can be found in Appendix~\ref{appendix:eval-results-details}.

\myparatight{Influence of $\genmodel$} Synthetic and natural data typically differ distributionally, even without watermarking. When clean clients fine-tune on synthetic data, distortion-free $\ITS$ and low-distortion $\Kirchenbauer$ with $\logitsbias=0$ introduce no bias relative to clean data. Table~\ref{tab:generative_robust} shows that $\randeigen$ fails to distinguish clean from watermarking clients in such cases ($\evasionrate \geq 60.2\%$). In contrast, when clean clients fine-tune on natural data, $\randeigen$ filters watermarking clients effectively ($\evasionrate \leq 0.7\%$). Since this shift gives the server an advantage, we stress-test $\Kirchenbauer$ by isolating $\genmodel$'s influence (i.e., having all clients fine-tune on synthetic data). $\Kirchenbauer$ remains non-robust: at $\temperature =0$ and $\logitsbias = 3$, its $\pval$ increases from a significant $1.01 \times 10^{-18}$ under $\vanillaFL$ to a non-significant 0.274 under $\activeFL$. Similarly, at $\temperature = 0.8$ and $\logitsbias = 3$, the $\pval$ increases from $1.03 \times 10^{-6}$ under $\vanillaFL$ to $0.157$ under $\activeFL$.

\myparatight{Influence of $\logitsbias$ and $\temperdiff$} Consistent with prior analysis~\citep{kirchenbauer2023watermark}, $\Kirchenbauer$ radioactivity improves with higher $\logitsbias$. Figure~\ref{tab:delta_radioactive} shows that $\pval$ drops from $10^{-3}$ to $10^{-6}$ as $\logitsbias$ increases from 3.0 to 6.0. Conversely, robustness degrades with higher $\logitsbias$ and $\temperdiff$. Table~\ref{tab:delta_robust} shows that $\evasionrate$ falls from 60.2\% to 0.7\% as $\logitsbias$ increases from $0$ to $5$. This is expected, as a larger $\logitsbias$ increases the watermark's deviation from the clean data distribution. $\Kirchenbauer$ robustness also degrades with larger $\temperdiff$. See Table~\ref{tab:temperature_robust} in Appendix~\ref{appendix sub:temperature} for details. 

\begin{embox}
\textbf{(RQ3)} 
Distributional shift stemming from  $\genmodel$ reduces watermark robustness. Larger $\logitsbias$ enhances $\Kirchenbauer$ radioactivity but reduces robustness. Larger $\temperdiff$ reduces $\Kirchenbauer$ robustness.
\end{embox}

\begin{table}[t]
\centering
\subcaptionbox*{}
{\begin{minipage}[t]{0.38\textwidth} 
    \centering
    \vspace{0pt}
    \setlength{\tabcolsep}{5pt}
    \caption{
Filtering is less effective when all clients use synthetic data ($\evasionrate \geq 60.2\%$), eliminating the shift introduced by $\genmodel$. In either case, $\Kirchenbauer$ is not radioactive.
}
\label{tab:generative_robust}
\centering
\footnotesize
\renewcommand{\arraystretch}{0.92}
\begin{tabular}{cccc}
\toprule
\textbf{WM} & \textbf{Clean Client}  & \textbf{$\bm{\evasionrate}$} & \textbf{$\bm{\overfilteringrate}$} \\
& \textbf{Dataset} & (\%) & (\%) \\
\midrule
$\Kirchenbauer$     & synthetic             & 60.2 & 92.9   \\
                    & natural              & 0.7 & 43.7   \\
\midrule
$\ITS$              & synthetic             & 60.7 & 92.9  \\
                    & natural              & 0.2 & 46.2  \\
\bottomrule
\end{tabular}

\end{minipage}}
\hfill
\subcaptionbox*{}
{\begin{minipage}[t]{0.56\textwidth}
    \centering
    \vspace{0pt}
    \setlength{\tabcolsep}{4.75pt}
    \footnotesize
\renewcommand{\arraystretch}{0.92}
\caption{
LLM watermark robustness with varying $\logitsbias$. 
Higher $\logitsbias$ makes filtering more effective (lower $\evasionrate$), decreasing watermark robustness.}
\label{tab:delta_robust}
\centering
\begin{tabular}{cccccc}
\toprule
\textbf{WM} & \textbf{$\bm{\logitsbias}$} & \multicolumn{2}{c}{\textbf{$\bm{p}$-value}} & \textbf{ER} & \textbf{OFR} \\
\cmidrule(lr){3-4} 
& & \textbf{Vanilla} & \textbf{Active} & (\%) & (\%) \\
\midrule
$\Kirchenbauer$ & 0 & 0.268 & 0.201 & 60.2 & 92.9  \\
                & 1 & $1.08 \times 10^{-5}$ & 0.370 & 21.6 & 80.6 \\
                & 3 & $1.01 \times 10^{-18}$ & 0.274 & 1.0 & 46.2 \\
                & 5 & $3.36 \times 10^{-12}$ & 0.788 & 0.7 & 40.7 \\
\midrule
$\ITS$   & -- & 0.480 & 0.480 & 60.7 & 92.9 \\
\bottomrule
\end{tabular}

\end{minipage}}
\vspace{-34pt}
\end{table}

\subsection{Utility Results}
\label{sec:eval-utility}

\myparatight{Influence of $\randeigen$} Table~\ref{tab:utility_2} and~\ref{tab:utility_3} reports the evaluation loss of the final global models from the experiments in Tables~\ref{tab:baseline_robust} and~\ref{tab:eps_robust}. The results show that server employing $\randeigen$ has a negligible effect on model utility: in the worst case, evaluation loss increases from 1.4818 under $\vanillaFL$ to 1.4868 under $\activeFL$, a relative increase of 0.34\%. This finding is consistent with \citet{lee2025practical}, who report an accuracy drop of less than 2\% with $\randeigen$ in their evaluated setting.

\myparatight{Influence of $\wmratio$} Although larger $\wmratio$ improves both radioactivity and robustness, it degrades utility. Evaluation loss rises from 3.156 to 3.161 as $\wmratio$ increases from 0\% to 30.0\%. Other benchmarks show a similar trend, discussed in Appendix~\ref{appendix:eval-results-details}. Thus, while $\wmratio$ can resolve the trade-off between radioactivity and robustness, it introduces a new, critical trade-off between watermark effectiveness and model performance.

\begin{embox}
\textbf{(RQ4)} Global model performance degrades with larger $\wmratio$, demonstrating a trade-off between model utility and watermark effectiveness.
\end{embox}

\begin{table}[t]
\centering
\subcaptionbox*{}
{\begin{minipage}[t]{0.50\textwidth}
    \centering
    \setlength{\tabcolsep}{5pt}
    \caption{Final evaluation loss for global model under $\vanillaFL$ and $\activeFL$ at $\wmratio=6.6\%$, corresponding to experiments in Table~\ref{tab:baseline_robust}. The largest observed loss increase is 0.34\%.}
\label{tab:utility_2}
\centering
\footnotesize
\renewcommand{\arraystretch}{0.92}
\begin{tabular}{cccc S[table-format=+1.2, explicit-sign=+]}
\toprule
\textbf{Data} & \textbf{Model} & \multicolumn{2}{c}{\textbf{Loss $\downarrow$}} & {\textbf{Rel. $\Delta$}} \\
\cmidrule(lr){3-4}
& & \textbf{Vanilla} & \textbf{Active} & {\textbf{(\%)}} \\
\midrule
$\cfour$    & 160M & 3.1571 & 3.1573 & 0.0063 \\
            & 410M & 2.8024 & 2.8026 & 0.0071 \\
\midrule
$\alpaca$   & 160M & 1.4818 & 1.4868 & 0.3374 \\
            & 410M & 1.2917 & 1.2542 & -2.9032 \\
\bottomrule
\end{tabular}
\end{minipage}}
\hfill
\subcaptionbox*{}
{\begin{minipage}[t]{0.46\textwidth} 
    \centering
    \setlength{\tabcolsep}{5pt}
    \caption{Final evaluation loss for global model under $\vanillaFL$ and $\activeFL$ with varying $\wmratio$, corresponding to Table~\ref{tab:eps_robust}. $\randeigen$ causes negligible utility degradation.}
\label{tab:utility_3}
\centering
\footnotesize
\renewcommand{\arraystretch}{0.92}
\begin{tabular}{ccc S[table-format=+1.2, explicit-sign=+]}
\toprule
\textbf{$\bm{\wmratio}$} & \multicolumn{2}{c}{\textbf{Loss $\downarrow$}} & {\textbf{Rel. $\Delta$}} \\
\cmidrule(lr){2-3}
& \textbf{Vanilla} & \textbf{Active} & {\textbf{(\%)}} \\
\midrule
6.6\%   & 3.1571 & 3.1573 & 0.0063 \\
16.6\%  & 3.1582 & 3.1592 & 0.0317 \\
30.0\%  & 3.1605 & 3.1621 & 0.0506 \\
\bottomrule
\end{tabular}
\end{minipage}}
\vspace{-34pt}
\end{table}

\section{Related Work}
Existing LLM watermarks fall into hashing-based schemes~\citep{kirchenbauer2023watermark, aaronson2023watermarking, christ2024undetectable} and non-hashing schemes~\citep{kuditipudi2023robust, zhao2023provable}. \citet{sander2024watermarking} show that hashing-based watermarks are radioactive in the centralized setting. We extend this to FL~\citep{pmlr-v54-mcmahan17a} and introduce an active-adversary threat model with the state-of-the-art strong Byzantine-robust aggregator~\citep{lee2025practical}. For further details on related work, refer to Appendix~\ref{sec:relate-work}. 
\section{Conclusion}

We study federated data provenance for LLMs and show that existing watermarks are radioactive under benign averaging yet removable by an active adversarial server using strong robust aggregation. Our findings reveal tensions among radioactivity, robustness, and utility in FL watermarking. We hope to advance understanding of current limitations and inform better design for future.
\section{Limitation}
Our evaluation is limited to settings in which clean clients' data are independently and identically distributed (IID), a single model family, and a single strong robust aggregation method. Despite that, our results highlight the tension between watermark radioactivity and robustness, motivating further investigation across more diverse settings.

\begin{ack}
\paragraph{Funding.} This work was supported in part by the National Science Foundation (NSF) Cybersecurity Innovation for Cyberinfrastructure (CICI) program under Grant No. 2531010. Any opinions, findings, and conclusions or recommendations expressed in this material are those of the authors and do not necessarily reflect the views of the NSF.

\paragraph{Competing interests.} We declare no relevant financial relationships during the 36 months preceding submission.
\end{ack}

\clearpage

\bibliographystyle{plainnat}
\bibliography{paper}
%%%%%%%%%%%%%%%%%%%%%%%%%%%%%%%%%%%%%%%%%%%%%%%%%%%%%%%%%%%%
\appendix
\section{Ethics Statement}
\label{append:ethics}
Our study investigates data provenance in federated learning using watermarks on synthetic LLM-generated text and does not involve interventions with human subjects or the collection of personally identifiable information. All experiments use public corpora $\cfour$~\citep{raffel2020exploring} and instruction datasets $\alpaca$~\citep{alpaca-stanford}, plus synthetic generations produced offline for participating clients. We explicitly model an active server as adversary to surface risks such that robust aggregation can be misused to suppress provenance signals, which could undermine transparency and attribution. We therefore discuss this threat model, quantify detectability/robustness trade-offs, and avoid releasing any tool meant to remove third-party watermarks. 

\section{Reproducibility}
\label{append:reproducibility}
The source code is available at \url{https://github.com/Caroline-111/FL-Watermark-Radioactivity}. The repository documents the FL protocol and watermarking schemes and includes scripts to (i) download and prepare datasets; (ii) generate watermarked text using $\Kirchenbauer$ and $\Kuditipudi$ with fixed random seeds; (iii) run $\vanillaFL$ and $\activeFL$; and (iv) compute all metrics from logged updates and summaries. We also provide the exact hyperparameters and random seeds to support reproduction of the reported results within the expected stochastic variation.

\section{The Use of Large Language Models}
\label{append:LLM}
In preparing this manuscript, we made limited use of a large language model solely for the purpose of polishing the writing. Specifically, the LLM was used to improve grammar, clarity, and readability of sentences drafted by the authors. The LLM did not contribute to research ideation, experimental design, analysis, or the generation of novel content. All scientific contributions, methodology, and results reported in this paper are the sole responsibility of the authors.

\section{LLM Watermarking}
\label{appendix:llm-wm-details}

We begin by formally defining LLM watermark radioactivity (Definition~\ref{def:radioactivity},~\citep{sander2024watermarking}), using the notation introduced in Section~\ref{sec:background-wm}.

\vspace{-1pt}
\begin{definition}[Radioactivity]
\label{def:radioactivity}
Dataset $\wmdata$ is $\signif$-radioactive for a statistical test $T$ with $\nullhypo$: Model $\globalM{}$ was not trained on $\wmdata$, if the test $T$ can reject $\nullhypo$ at a $p$-value below the significance level $\signif$.
\end{definition}

We adapt \Kirchenbauer\ and \Kuditipudi\ to the radioactivity setting by applying the cumulative scores (\Kirchenbauer) or alignment-based statistics (\Kuditipudi) to model predictions on $\wmdata$. We construct the corresponding null distributions using models that were not trained on $\wmdata$. The remainder of this section details the generation and detection functions used by these two schemes.

\subsection{\citep{kirchenbauer2023watermark}}
\Kirchenbauer\ is a hashing-based {\em green list} watermark. At step $t$, the scheme hashed the previous $k$ tokens together with a secret key $s$ to seed a PRNG that randomly partitions the vocabulary $V$ into a green list $G_t$ of size $\gamma|V|$ and a redlist $R_t$. The logits of green tokens are shifted by $\delta > 0$ before softmax. Sampling then proceeds from the biased distribution $\hat{p}^{(t)}$. Parameters $\gamma\in(0,1)$ and $\delta$ determine the trade-off of strength and quality of the watermark.

Given a tokenized text of $N$ tuples $X_N=\{x_i\}_{i=1}^N$, define the per-token indicator
$$
\text{WScore}(x^{(0)}_i;\, s, x^{(-1)}_i,\ldots,x^{(-k)}_i)\;=\;\mathbf{1}\{x^{(0)}_i\in G_i\}.
$$

The cumulative score $S(X_N)=\sum_{i=1}^N \text{WScore}(\cdot)$ counts green tokens. Under $H_0$ where no watermark exists, the indicators are i.i.d. $\mathrm{Bernoulli}(\gamma)$, in this case $S\sim \mathrm{Binomial}(N,\gamma)$ and a one-sided $p$-value follows from the binomial CDF. A $z$-test form is
$$
z=\frac{S-\gamma N}{\sqrt{N\gamma(1-\gamma)}},
$$
with small one-sided $p\text{-value}$ indicating detection. The watermark functioning by logit boost $\delta$ on $G_t$ preserves quality in low-entropy context, while still yielding predictable sensitivity curves as N grows.

\subsection{\citet{kuditipudi2023robust}}
\Kuditipudi\ is a distortion-free, key-sequence watermark. A shared random key sequence $\xi=(\xi_1,\dots,\xi_n)$ drives a decoder $\Gamma$ that maps $(\xi_i,\,p^{(t)})$ to a next token while preserving the model's original sampling distribution, marginalizing over $\xi$. The watermark proposes two mechanisms:
\begin{itemize}
    \item $\ITS$ (inverse-transform sampling): $\Gamma$ uses a uniform $u\in[0,1]$ and a random permutation $\pi$ to pick the first token whose CDF which is ordered by $\pi$ exceeds $u$. This decoder is provably distortion-free.
    \item $\EXP$ (exponential-minimum, Gumbel-style): $\Gamma$ draws i.i.d. exponential variables keyed by $\xi$ and selects $\arg\min_i E_i/p_i$, yielding an equivalent distortion-free sample. In practice it also uses a shift-genenrate wrapper that slices fresh subsequences of $\xi$ to avoid reuse while retaining detectability.
\end{itemize}

The detection of \Kuditipudi\ watermark is model- and prompt-agnostic. Given candidate text $y$ and shared key $\xi$, \Kuditipudi\ aligns $y$ to $\xi$ using an alignment cost $d$ and then computes a nonparametric permutation test $p$-value by comparing the observed cost to costs under resampled keys $\{\xi^{(t)}\}$:

$$
\hat p=\frac{1+\sum_{t=1}^T \mathbf{1}\{\,\phi(y,\xi^{(t)})\le \phi(y,\xi)\,\}}{T+1},\quad \phi(y,\xi)=\min_{\text{blocks}} d(\text{block}(y),\text{block}(\xi)).
$$

where $d$ is the alignment cost. A simple and fast alignment cost used in ITS is

$$
d\bigl(y,(u,\pi)\bigr)=\sum_{i=1}^{\text{len}(y)}\bigl|u_i-\eta(\pi(y_i))\bigr|, \quad \eta(j)=\tfrac{j-1}{|V|-1}.
$$

For edit robustness, \Kuditipudi\ employs a Levenshtein-style variant $d_\gamma$ that permits insertions or deletions with penalty $\gamma$. Under $H_0$ where key is independent of the text, the permutation test is valid and yields calibrated $p$-values. Statistical power grows exponentially in text length and only linearly degrades with key length.

\section{Byzantine Robust Aggregation}
\label{appendix:Byzantine-robust-agg}

The aggregator $\fil$ takes a set $\allupdates$ of $N$ vectors in $\mathbb{R}^d$, of which an $\wmratio$-fraction may be arbitrarily corrupted, as inputs. Let $\cleanupdates \subseteq \allupdates$ denote the subset of uncorrupted vectors, with mean
\begin{align*}
\mu_C
= \frac{1}{|\cleanupdates|}
\sum_{\updates{i}{t} \in \cleanupdates}
\updates{i}{t}.
\end{align*}
$\fil$ guarantees:
\begin{align*}
    bias = \left\| \fil(\allupdates) - \mu_C \right\|_2 \leq \beta \cdot \|\Sigma_C\|_2^{\frac{1}{2}}
\end{align*}
where $\|\Sigma_C\|_2$ is the spectral norm of the covariance matrix of $\cleanupdates$. Robust aggregators security is closely related to the multiplicative factor $\beta$: a smaller $\beta$ corresponds to a tighter defense. This factor distinguishes strongly-bounded aggregators from weakly-bounded ones. Weak robust aggregators have $\beta = O(d^{\frac{1}{2}})$, while strong robust aggregators have $\beta = O(1)$, independent of the vector dimension $d$~\citep{diakonikolas2017being}.

\subsection{Polynomial-time Strong Robust Aggregators}

\begin{algorithm}
	\caption{Meta-Algorithm for Strong Byzantine Robust Aggregators} 
    \label{alg:aggregator}
    \hspace*{\algorithmicindent} \textbf{Input} Watermark ratio $\wmratio$, partially watermarked updates $\allupdates = \{\updates{1}{t}, \cdots, \updates{N}{t}\} \subseteq \mathbb{R}^d$,  upper bound of clean covariance $\cleansigma$ as $\Gamma$ 
    \\
    \hspace*{\algorithmicindent} \textbf{Output} Robust aggregated mean $\mu$
	\begin{algorithmic}[1]
        \For{$j = 0, \cdots, 2 \cdot \wmratio  \cdot N- 1$} \label{aggregator:line:1} 
            \State Compute current maximum eigenvalue $\lambda_{curr} = \|Cov(\allupdates) \|_2$
            \If{$\lambda_{curr} \leq \Gamma$} \label{aggregator:line:3} 
                break
            \Else
                \State $\allupdates \leftarrow \outlierremoval(\allupdates, \wmratio, \cleansigma)$
            \EndIf
        \EndFor
        \State \Return $\mu = \frac{1}{|\allupdates|} \sum_{\updates{i}{t} \in \allupdates} \updates{i}{t}$
	\end{algorithmic} 
\end{algorithm}
 
Algorithm~\ref{alg:aggregator} shows a meta-algorithm for polynomial-time \textit{strong} Byzantine robust aggregation functions. Polynomial-time strong robust aggregators~\citep{diakonikolas2017being, hopkins2020robust, kothari2017outlier} share a common strategy: they iteratively attenuate outliers in $X$ with an $\outlierremoval$ until the bias is provably bounded. This subroutine, whose implementation varies by schemes, is designed to filter or down-weight outliers that contribute most to $X$ variance along the direction of largest eigenvectors. The iterative filtering continues until \circled{1} Number of iterations reaches $2\wmratio N$ (line~\ref{aggregator:line:1}), assuming at least one point is removed per round or \circled{2} largest eigenvalue of the current set's covariance matrix falls below a predefined threshold $\Gamma$ (line~\ref{aggregator:line:3}). 

The first condition is derived from the proof that removing at least $2\wmratio N$ vectors is sufficient to eliminate all malicious vectors. The second condition originates from the observation that in practice the bias is bounded once the largest eigenvalue ($\lambda_{curr}$) falls below a predetermined threshold $\Gamma$, where $\Gamma = k \|\Sigma_Y\|_2$ for $k \in [\sqrt{20}, 9]$~\citep{pmlr-v206-zhu23b, diakonikolas2019robust, diakonikolas2019recent}. Note that these methods require $O(\epsilon d^2)$ operations and $O(\ d^2)$ memory, making them impractical and vulnerable to attacks in high-dimensional settings~\citep{choudhary2024attacking, lee2025practical}. 

\subsection{Quasi-Linear Strong Robust Aggregators}
We employ $\randeigen$~\citep{lee2025practical} in our $\activeFL$. The strong Byzantine robust aggregator runs in quasi-linear running time, $O(Nd)$, and possesses provably near-optimal bias bounds. Unlike polynomial-time strong aggregators, it does not require computing the maximum variance of clean vectors, $\|\Sigma_C\|_2$. Instead, they replace the stopping condition in line~\ref{aggregator:line:3} from Algorithm~\ref{alg:aggregator} with an equivalent heuristic based on eigenvalue convergence: the algorithm terminates once the maximum variance of the iteratively filtered set stabilizes. To avoid computational bottleneck, $\randeigen$ also estimate the dominant eigenvectors through randomized dimensionality reduction rather than computing them exactly.

\section{Problem Statement}
\label{appendix:ps}
This section provides additional details on the threat model and formally defines watermark robustness in FL.

\subsection{Threat Model}
For simplicity, clients have local datasets of equal size, so no weighting across clients is required. If none of the clients apply watermarking, the data is independent and identically distributed. We have two types of clients. Watermarking clients are assumed to constitute $\epsilon$ of the total clients. All watermarking clients collude, sharing the same watermarking technique and random seed, without sharing their clean local data.
Clean clients are local clients that compute updates using clean datasets without any watermarking. They operate independently, without communication or collaboration with watermark clients or among themselves. Each clean client has access only to its own local training data, which neither clients nor server cannot access. The central server is the active adversary. It controls the aggregation process of local updates, where the simplest method for aggregating is averaging~\citep{pmlr-v54-mcmahan17a}. The server is free to choose the aggregation function. Its goal is to obtain a model that evades detection of the watermark while maintaining the global model performance.

\subsection{Watermark Robustness Definition}
The adversary in \activeFL aims to $\circled{1}$ obtain an updated model $\globalM{t+1}$ that has a similar utility to $\vanilla$ on a set of clean updates under some evaluation metric $\mathcal{E}$\footnote{Evaluation metrics include log-likelihood on a test set or  qualitative metrics such as BLEU~\citep{bleuscore} or ROUGE~\citep{lin-2004-rouge}.} 
and $\circled{2}$ reduce the detectability of the watermark. Specifically, it aims to reduce the statistical significance $\alpha$ of a given watermarked dataset $\wmdata_i$ for $\globalM{t+1}$. We use $\detect{\globalM{t+1},\vanilla}{\secretkey}(\wmdata_i)$ to denote that the detection test is run on the predictions of $\globalM{t+1}$ on the watermarked dataset $\wmdata_i$ at round $t$, where $\globalM{t+1}$ is obtained by training with $\vanilla$. We use $\approx_\mathcal{E}$ to denote similar utility under the metric $\mathcal{E}$. Note that radioactivity (Definition~\ref{def:radioactivity}) is defined for a given dataset $\wmdata$, not a watermarking scheme. We thus also define {\em robustness} with respect to $\adversary$ and to a dataset that is $\alpha$-radioactive had it been updated with $\vanilla$.  

\begin{definition}[FL Robustness]
\label{def:robustness}
Let $\wmdata_i$ be an $\alpha$-radioactive dataset for a statistical test $T$ and the model $\globalM{t+1}$ obtained in \vanillaFL with $\vanilla$ such that $\detect{\globalM{t+1},\vanilla}{\secretkey}(\wmdata_i) \returns \true$ at round $t$.
If there exists an adversary $\adversary$ such that for every round $t_\adversary$: $\circled{1}~\adversary(\allupdates,\globalparams{t_\adversary}) \approx_\mathcal{E} \vanilla(\cleanupdates,\globalparams{t_\adversary})$ and $\circled{2}~\detect{\globalM{t_\adversary+1},\adversary}{\secretkey}(\wmdata_i) \returns \false$, then $\wmdata_i$ is not robust to $\adversary$.
\end{definition}

The robustness definition is {\em counterfactual}, similar to its counterpart for watermarked generations of an LLM~\citep{zhao2025sok}. 
It has a hypothetical precondition that at some round $t$, the dataset $\wmdata_i$ would become radioactive on the resulting global model $\globalM{t+1}$, if it had been updated with $\vanilla$.
Given that $\wmdata_i$ satisfies this precondition, if the server runs $\adversary$ for each round $t_\adversary$ and the dataset $\wmdata_i$ is not radioactive on the resulting model $\globalM{t_\adversary+1}$, then we say that the $\wmdata_i$ is not robust to $\adversary$.
On the contrary, a watermarked dataset $\wmdata_i$ is robust to $\adversary$ if it remains detectable in spite of  $\adversary$ updating the model for all rounds $t_\adversary$, $\adversary(\allupdates,\globalparams{t_\adversary})$.
While increasing the rounds $t$ means that the dataset's radioactivity increases (lower $p$-value), it comes at a cost of overfitting which can damage the final model utility.
If the datasets are not radioactive to start with, then the server trivially satisfies condition $\circled{2}$. If the server does not satisfy condition $\circled{1}$, then condition $\circled{2}$ is easy to satisfy. For instance, adding noise to all updates could sufficiently perturb the watermark signal but such a strategy can deteriorate the final $\globalM{}$ utility. 

\section{Data-Provenance Framework for Federated Fine-Tuning}
\label{appendix:imple-setup-details}

\begin{algorithm}
    \caption{LLM Watermarks in Federated Learning Finetuning} 
    \label{alg:wm_in_fl}
    \hspace*{\algorithmicindent} \textbf{Input} Clients $\setclients$, local datasets $\{\traindata_i\}_{i=1}^\allworker$,  initial global model parameters $\globalparams{0}$, server learning rate $\eta_s$, client learning rate $\eta_c$, local training steps $J$
	\begin{algorithmic}[1]
        \For{each client $i \in \mathcal{C}$}
            \If{client $i$ chooses to apply watermark}
                \State $\wmdata_i = \watermark{\genmodel}{\secretkey}(\traindata_i)$
                \State $\traindata_i \gets \wmdata_i$ \Comment{Client $i$ uses $\wmdata_i$ as local dataset}
            \EndIf
        \EndFor
        \State Initialize $t = 0$
		\While{validation loss decreases} \Comment{Early stopping}
            \State $\localparams{i, 0}{t} = \globalparams{t}$       \label{wm_in_fl:line:7} \Comment{Global model broadcast parameters to all clients}
			\For{each client $i \in \mathcal{C}$ \textbf{in parallel}}

                \State Train on $\traindata_i$ with $\clientopt$ and $\eta_c$ for $J$ steps
                \State $\updates{i}{t} = \localparams{i, J}{t} - \globalparams{t}$ \Comment{Compute local updates}
			\EndFor
            \State $\updates{}{t} = \agg (\{ \updates{i}{t}\}_{i=1}^\allworker)$ \label{wm_in_fl:line:13} \Comment{The server employs $\fil$ as $\agg$ in $\activeFL$}
            \State $\globalparams{t+1} = \serveropt(\globalparams{t}, -\updates{}{t}, \eta_s)$ \Comment{Update global model}
            \State $t \gets t + 1$
        \EndWhile
        \State Each client $i$ runs $\detect{\globalM{t}}{\secretkey}(\wmdata_i)$
	\end{algorithmic} 
\end{algorithm}

In this section, we present Algorithm~\ref{alg:wm_in_fl}, which implements the data-provenance framework for FL fine-tuning introduced in Section~\ref{sec:approach}. Our FL framework builds upon \FedOpt~\citep{reddi2020adaptive}, incorporating two key modifications. First, to accelerate convergence, we broadcast the global model to all clients in each round $t$ (line~\ref{wm_in_fl:line:7}), rather than to a subset as in \FedOpt. In parallel, each client then performs local training on its own dataset using $\clientopt$ with learning rate $\eta_c$ and sends the resulting model updates to the server. 

The second adaptation is the aggregation step. The central server aggregates the received updates using a function $\agg$ (line~\ref{wm_in_fl:line:13}). $\FedOpt$ is a specialized case where $\agg$ is simply averaging, which we adopt as $\vanillaFL$. In $\activeFL$, $\agg$ is instead a strong Byzantine robust aggregation function. The server then updates the global model by applying $\serveropt$ with learning rate $\eta_s$ on the aggregated results. Upon completion, the clients perform watermark detection on the final global model $\globalM{t}$ using their own watermarked datasets.

\section{Experimental Setup}
\label{subsec:imple-setup-details}

\subsection{FL Setup}
We adopt the FL framework from FedOPT~\citep{reddi2020adaptive} to achieve faster convergence with good accuracy. Specifically, we employ ADAM~\citep{kingma2014adam} as the server optimizer, which maintains optimizer states across communication rounds, and employ normalized SGD as the client optimizer. A constant learning rate of $10^{-5}$ is used for both optimizers. During each communication round, local fine-tuning is performed for a single epoch on the respective client datasets. We employ an early stopping strategy that terminates training once the evaluation loss on a clean, held-out validation split fails to decrease for three consecutive rounds. 

\subsection{LLM Watermarks Setup}
We evaluate two representative LLM watermarking schemes: $\Kirchenbauer$~\citep{kirchenbauer2023watermark} and $\Kuditipudi$~\citep{kuditipudi2023robust}. $\Kuditipudi$ proposes two mechanisms, $\ITS$ and $\EXP$. In this paper, we evaluate $\ITS$.

We employ pretrained Pythia-2.8B as our generative model ($\genmodel$) to ensure high-quality text generation and avoid potential biases that could arise from using the same models for $\globalM{}$ and $\genmodel$ \citep{sander2024watermarking}. On the question-answer pairs dataset $\alpaca$~\citep{alpaca-stanford}, we follow the original implementation to generate watermarked responses~\citep{kirchenbauer2023watermark}. On the raw-text dataset $\cfour$~\citep{raffel2020exploring}, we treat the first 20 tokens as a prompt and generate the watermarked content to the same length as the original text. For the $\Kirchenbauer$ baseline experiments, we use a softmax temperature of 0.8 and a logit bias of 3.0 following prior work's experimental setup \citep{sander2024watermarking}. This configuration ensures watermark detectability without compromising text quality. For radioactivity detection, we adopt the setting of open model and supervised training data access, proposed by~\citet{sander2024watermarking}. For the $\Kuditipudi$ baseline experiments, we sample each next token from the model’s full softmax distribution with $\temperature=1$ and no top-k or top-p truncation. Each dataset sample is paired with a pre-generated key sequence of the same length.

\subsection{Dataset Setup}
All clients have local datasets of the same size (i.e., the same number of prompts). In our baseline experiments, before any clients apply watermarks, we assume that all data are natural, non-watermarked, and IID across clients. When certain clients opt to watermark their dataset, they generate watermarked responses to all prompts from their original natural datasets, creating a watermarked, synthetic dataset. In the ablation study for watermark robustness, we examine the setup where the clean clients also use the same generative model to generate non-watermarked, synthetic dataset. The same evaluation conclusions hold for clean clients fine-tune on either natural or synthetic dataset. 

To stress-test watermark robustness, we vary the decoding strategy for generating non-watermarked synthetic datasets based on the watermarking scheme. 
\begin{enumerate}
    \item \textbf{\citet{kuditipudi2023robust}}: Since $\Kuditipudi$ does not employ a softmax temperature ($\temperature$), we use standard Multinomial Sampling ($\temperature = 1$) to generate non-watermarked dataset 
    \item \textbf{~\citet{kirchenbauer2023watermark}}:  $\Kirchenbauer$ uses $\temperature$ as a hyperparameter. We use a greedy decoding strategy ($\temperature = 0$) to generate non-watermarked dataset. To create a softmax temperature difference ($\temperdiff$) between watermarked and clean datasets, we set the watermarking scheme's $\temperature$ to exactly $\temperdiff$. 
\end{enumerate}

\subsection{Evaluation Metrics Setup}
We evaluate watermark robustness by computing two metrics for each layer and reporting their averages across layers. Here, a \emph{layer} refers to an individual parameterized module in LLM, such as a multi-head self-attention layer. Let $\numlayer$ be the total number of layers. For each layer $\eachlayer$, let $\wmworkerl$ be the set of watermarking clients and $\filterworkerl$ be the set of all clients filtered by the aggregator. Averaging across all $\numlayer$ layers, we define 1) \textbf{Evasion Rate ($\bm{\evasionrate}$)} as the mean fraction of watermarking clients that remain after aggregation: $\evasionrate = \avgfilter (1 - \frac{|\wmfilterworkerl|}{|\wmworkerl|})$; and 2) 
\textbf{Overfiltering Rate ($\bm{\overfilteringrate}$)} as the mean fraction of filtered clients that are not watermarked: $\overfilteringrate = \avgfilter (1 - \frac{|\wmfilterworkerl|}{|\filterworkerl|})$.

\section{Evaluation Results}
\label{appendix:eval-results-details}

\begin{table}[t]
\caption{LLM watermarks robustness with varying $\temperdiff$. Higher $\temperdiff$ increases distributional shift, making filtering more effective (lower $\evasionrate$).}
\label{tab:temperature_robust}
\begin{center}
\small
\begin{tabular}{cccccc}
\toprule
\textbf{WM} & \textbf{$\bm{\temperdiff}$} & \textbf{$\bm{\evasionrate}$(\%)} & \textbf{$\bm{\overfilteringrate}$(\%)} \\
\midrule

$\Kirchenbauer$ & 0.0 & 60.2 & 92.9   \\
                & 0.4 & 8.9 & 66.7 \\
                & 0.8 & 1.8 & 44.3 \\
\midrule
$\Kuditipudi$   & --  & 60.7 & 92.9  \\
\bottomrule
\end{tabular}
\end{center}
\end{table}

\begin{table}[t]
\caption{Per-Client Detection Results ($p$-values)}
\label{tab:individual_detect}
\centering
\small
\begin{tabular}{ccccccccc}
\toprule
\textbf{Data} & \textbf{WM} & \textbf{Model} & \multicolumn{2}{c}{\textbf{Client 1}} & \multicolumn{2}{c}{\textbf{Client 2}}\\
\cmidrule(lr){4-7}
& & & \textbf{Before FT} & \textbf{After FT} & \textbf{Before FT} & \textbf{After FT} \\
\midrule
$\cfour$    &   $\Kirchenbauer$ & 160M & 0.390 & $4 \times 10^{-3}$ & 0.447 & $4\times 10^{-2}$\\
            &                   & 410M & 0.839 & $4.55 \times 10^{-5}$ & 0.725 & $5.84 \times 10^{-5}$\\
\bottomrule
\end{tabular}
\end{table}

\subsection{Watermark Hyperparameters Analysis}
To evaluate how $\logitsbias$ affects watermark radioactivity, we use the same setting as the baseline Pythia-160M experiment, varying only $\logitsbias$. To evaluate watermark robustness, we test three hyperparameters: the generative model ($\genmodel$), logits bias ($\logitsbias$), and difference in softmax temperature used to generate clean and watermarked dataset ($\temperdiff$). For each experiment, we vary one parameter while limiting the influence of the others. When limiting the influence of $\genmodel$, clean clients fine-tune on synthetic datasets. To limit the influence of $\logitsbias$ and $\temperdiff$, we set them at $0$.

\subsubsection{Influence of $\logitsbias$} As stated in Section~\ref{subsec:hyperparameters}, $\Kirchenbauer$ radioactivity improves with a higher $\logitsbias$. However, this trend only holds within a specific range. An excessively low $\logitsbias$ induces high repetition in the watermarked text. Since the detection method is designed to ignore such repetitions~\citep{sander2024watermarking}, the effective number of tokens evaluated becomes small. This compromises the statistical power of the detection test, leading to artificially low $\pvals$ under $\nullhypo$ and a high false positive rate. Conversely, an excessively high $\logitsbias$ makes the watermarked text resemble a sequence of loosely connected tokens, because the added bias overwhelms the model's original prediction for the next token. Such sequences are more difficult for the global model to learn than meaningful, coherent sentences during fine-tuning, ultimately leading to higher $\pval$ after fine-tuning.

\subsubsection{Influence of Softmax $\temperdiff$}
\label{appendix sub:temperature}

$\Kirchenbauer$ robustness degrades with a higher $\temperdiff$. A larger $\temperdiff$ introduces a larger distributional shift, allowing $\randeigen$ to filter out watermarking clients more effectively. Table~\ref{tab:temperature_robust} shows that increasing $\temperdiff$ from 0.0 to 0.8 reduces $\evasionrate$ from $60.2\%$ to $1.8\%$ and $\overfilteringrate$ from $92.9\%$ to $44.3\%$. $\temperature$ is not a hyperparameter of $\Kuditipudi$. We only test the setting where the clean and watermark synthetic dataset share the same $\temperature$ ($\temperature=1$ and $\temperdiff = 0$), whose metrics are similar to those of $\Kirchenbauer$ at $\temperdiff = 0$.

\subsection{Individual Detection for Each Client}
\label{appendix sub:individual-detection}

For simplicity, detection is run on the aggregated watermarked dataset (across all clients). However, this methodological choice does not affect our conclusions. To confirm, we examine per-client detection in the worst-case scenario: We use the $\cfour$ dataset with 6.6\% $\Kirchenbauer$ watermark, which is radioactive with aggregated detection but has relatively high $\pvals$ compared to other configurations. Table~\ref{tab:individual_detect} shows that even in this worst-case scenario, per-client detection still yields high $\pvals$ before fine-tuning (around $0.5$) and statistically significantly low $\pval$ after fine-tuning ($\pval <= 0.04$).

\subsection{Overfitting Points}
In Table~\ref{tab:overfitting_rounds}, we report the exact number of communication rounds before overfitting occurs for the experiments presented in Table~\ref{tab:baseline_radioactive} and Table~\ref{tab:baseline_robust}. We observe that dataset $\cfour$ needs about $30$ rounds to converge, while dataset $\alpaca$ needs approximately $100$ rounds to converge. Compared to $\vanillaFL$, $\activeFL$ typically needs a few less rounds to converge. 

\begin{table}[t]
\caption{Communication rounds before overfitting occurs for varied FL configurations.}
\label{tab:overfitting_rounds}
\centering
\small
\begin{tabular}{ccccc}
\toprule
\textbf{Data} & \textbf{WM} & \textbf{Model} & \multicolumn{2}{c}{\textbf{Overfitting Round}}\\
\cmidrule(lr){4-5}
& & & \textbf{Vanilla} & \textbf{Active} \\
\midrule
$\cfour$    &   $\Kirchenbauer$ & 70M  & 31 & -- \\
            &                   & 160M & 37 & 35 \\
            &                   & 410M & 33 & 30 \\
\cmidrule(lr){2-5}
            &   $\Kuditipudi$   & 70M  & 31 & --  \\
            &                   & 160M & 38 & --  \\
            &                   & 410M & 29 & --  \\
\midrule
$\alpaca$   &   $\Kirchenbauer$ & 70M  & 97 & --  \\
            &                   & 160M & 124 & 113 \\
            &                   & 410M & 97 & 91  \\
\cmidrule(lr){2-5}
            &   $\Kuditipudi$   & 70M  & 104 & --  \\
            &                   & 160M & 136 & --  \\
            &                   & 410M & 99 & --  \\
\bottomrule
\end{tabular}
\end{table}

\subsection{Utility of the Global Model}
\label{sec:formal-def-detail}

\begin{definition}[Utility]
\label{def:generalization}
Let $\globalM{}(\cleandata)$ and $\globalM{}(\wmdata)$ be global models trained on a clean dataset $\cleandata$ and its watermarked version $\wmdata$, respectively. A watermarking scheme is $\lambda$-generalizable if, for any unseen dataset $\traindata$ and any bounded evaluation metric $\eval{\mathcal{M}}{\traindata}: \mathcal{M} \times \traindata^* \rightarrow R$, the following holds:
\begin{align}
\left| \eval{\globalM{}(\cleandata)}{\traindata} - \eval{\globalM{}(\wmdata)}{\traindata} \right| \leq \lambda.
\end{align}
\end{definition}

We evaluate the impact of watermark strength on downstream utility by varying the watermark ratio $\wmratio$ during training of 160M-Pythia, and assessing word-level perplexity on $\cfour$~\citep{raffel2020exploring} and question-answering accuracy on MMLU-Pro~\citep{wang2024mmlu}. As shown in Table~\ref{tab:performance_appendix}, increasing $\wmratio$ from 0\% to 16.6\% slightly degrades both metrics, with higher perplexity and lower accuracy. However, at $\wmratio=30\%$, the trend reverses, despite a marginally higher entropy loss of 3.1605, the model achieves its best C4 perplexity of 85.792 (vs.\ 87.029 at 0\%) and highest MMLU-Pro accuracy of 0.11827 (vs.\ 0.11760).

A plausible explanation is that a sufficiently strong watermark acts as a data-dependent regularizer. The green list bias more consistently guides learning toward on-manifold continuations, implicitly smoothing labels by redistributing probability mass among plausible tokens and reducing gradient variance from rare tail tokens. When $\wmratio$ is too low (e.g., 6.6–16.6\%), the bias perturbs the logits without providing these regularization benefits. At $\wmratio=30\%$, however, the bias appears to surpass a useful-strength threshold, where improved calibration and stability outweigh the small increase in token-level cross-entropy, resulting in better generalization on both language modeling and reasoning tasks.

\subsection{Other Watermarks}
\label{appendix:other}

\subsubsection{LLM Watermarks}
Other than $\Kirchenbauer$~\citep{kirchenbauer2023watermark} and $\Kuditipudi$~\citep{ kuditipudi2023robust}, we examine another representative LLM watermark, AAR~\citep{aaronson2023watermarking}, which is computationally distortion-free, provided hashing context (previous $k$ tokens, $k$-gram) never repeats~\citep{zhao2025sok}.
As a hashing-based method, AAR exhibits high radioactivity similar to $\Kirchenbauer$ in the FL setup~\citep{sander2024watermarking}. Similarly, it is also not robust against $\randeigen$. Under the same setup as the $\Kirchenbauer$ experiments ($k$-gram = 4 and all clients fine-tune on synthetic data), AAR yields $\evasionrate = 1.4\%$. This happens because natural language is highly repetitive, causing the hashing window to repeat often in practice. Once this distortion-free assumption breaks down, the watermark deviates from the clean update distribution, making it easy for $\randeigen$ to filter out watermarked updates. 

\subsubsection{Non-LLM Watermarks}
There are non-LLM watermarking schemes applicable to the FL fine-tuning, such as post-hoc detection. We examine the Unicode character replacement scheme~\citep{wei2024proving}, which introduces a simple watermark by inserting random sequences or Unicode lookalikes into training data. This approach, independent of the LLM generation process, detects watermarks by analyzing token loss over these modified elements. We find that this watermark method lacks sufficient detection significance under the FL setting. We test the watermark on Pythia-70M with $\wmratio=30\%$, leading to a $p$-value of $0.274$. In comparison, $p$-value for $\Kirchenbauer$ under the same setting of $\wmratio=30\%$ is $0.0041$. The likely reason for this insignificance is that the original method was designed for pre-training; FL fine-tuning appears inadequate for the watermark to become statistically discernible.

\begin{table}[t]
\caption{Evaluation of 160M-Pythia with varying $\Kirchenbauer$ watermark proportion ($\wmratio$).}
\label{tab:performance_appendix}
\begin{center}
\begin{tabular}{cccc}
\toprule
\textbf{$\bm{\wmratio}$} & \textbf{Entropy Loss $\downarrow$} & \makecell{\textbf{C4~\citep{raffel2020exploring}}\\Word Perplexity $\downarrow$} & \makecell{\textbf{MMLU-Pro~\citep{wang2024mmlu}}\\Accuracy $\uparrow$} \\
\midrule
0\%     & 3.1564 & 87.029 & 0.11760 \\
6.6\%   & 3.1571 & 87.349 & 0.11735 \\
16.6\%  & 3.1582 & 91.279 & 0.11652 \\
30.0\%  & 3.1605 & 85.792 & 0.11827 \\
\bottomrule
\end{tabular}
\end{center}
\end{table}

\section{Local Updates Analysis}
\label{appendix:updates-analysis}

\begin{figure*}[!t]%
    \centering
    \begin{subfigure}{0.48\textwidth}
    \includegraphics[width=\textwidth]{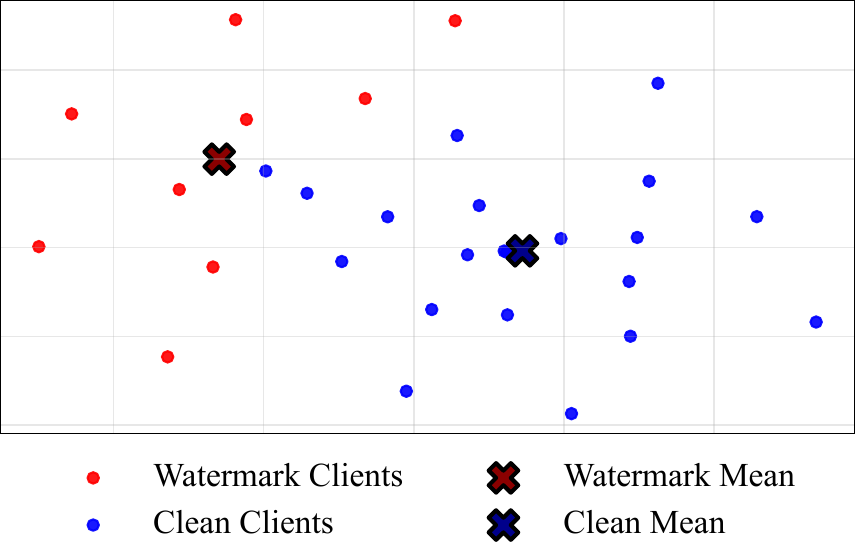}
    \caption{\Kirchenbauer}
    \label{fig:tsne_k_allgen}
    \end{subfigure}
    \quad
    \begin{subfigure}{0.48\textwidth}
    \includegraphics[width=\textwidth]{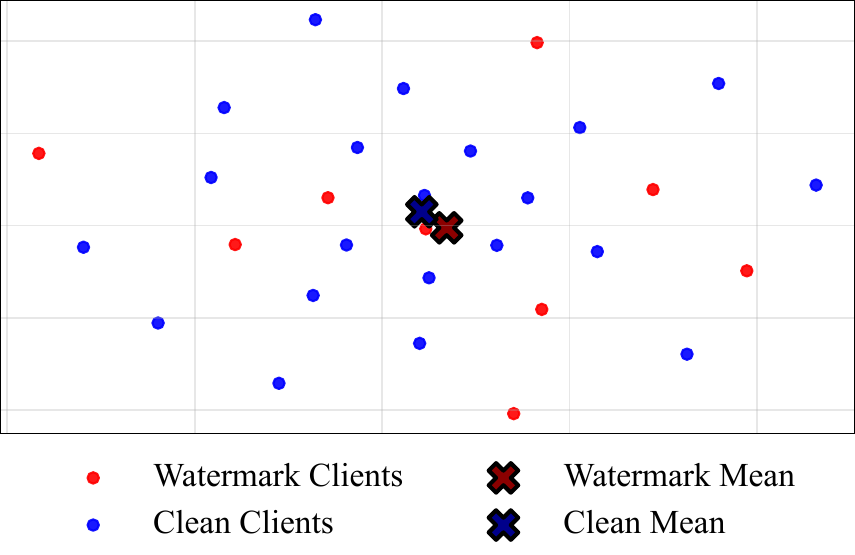}
    \caption{\Kuditipudi}
    \label{fig:tsne_p_allgen}
    \end{subfigure}
    \caption{t-SNE visualizations of local updates from clean clients (blue) and watermarking clients (red) for \Kirchenbauer\  and \Kuditipudi\ in the first round. The mean gradient for each group is marked with an 'X'. All clean clients use synthetic data.}
    \label{fig:gradient-diff-type}
\end{figure*}

From Figure~\ref{fig:gradient_distributions} where clean clients trained on raw \cfour\ data, we can observe clear and significant separation between the updates from clean clients and those from watermarking clients. This distinct separation strongly indicates that the watermarking data alters the model updates in a detectable manner, making the two groups of clients easily distinguishable in the high-dimensional space. We can also observe that the gradients from watermarking clients are similar to each other, forming a very tight and dense cluster. This highlights a high degree of local similarity and consistency among the watermark updates. This indicates that watermark's effect is consistent and pronounced and it could potentially be detected by using deviation from the robust mean estimation. Figures~\ref{fig:tsne_k_allgen} and \ref{fig:tsne_p_allgen} visualize the impact of watermarking on client updates when clean clients use synthetic data. Figure~\ref{fig:tsne_k_allgen} shows that watermark and clean updates still form two well-separated groups in high-dimensional space, though the two clusters are much closer than in Figure~\ref{fig:gradient_distributions}. In contrast, Figure~\ref{fig:tsne_p_allgen} shows that the \Kuditipudi\ watermark does not have obvious effect on client updates and clean and watermarking clients are not well separated in high-dimensional space. This result is consistent with the low radioactivity reported in Table~\ref{tab:baseline_radioactive}.

\begin{figure*}[!t]%
    \centering
    \begin{subfigure}{1\textwidth}
    \includegraphics[width=\textwidth]{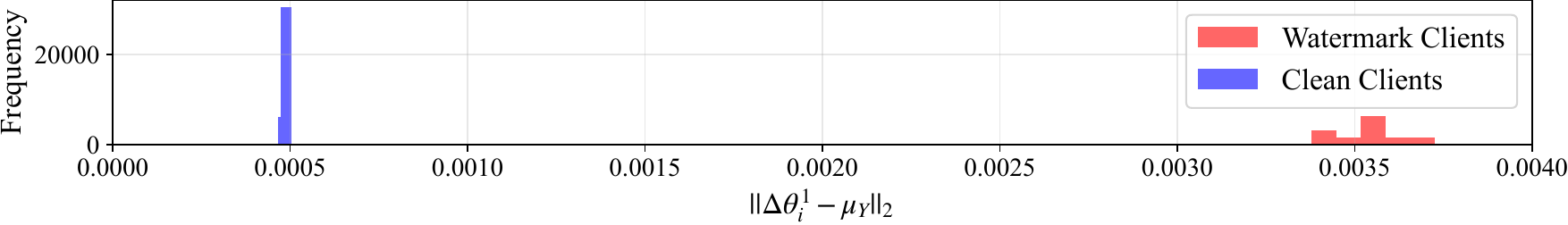}
    \caption{\Kirchenbauer}
    \label{fig:tsne_k_allgen_dist}
    \end{subfigure}
    \begin{subfigure}{1\textwidth}
    \includegraphics[width=\textwidth]{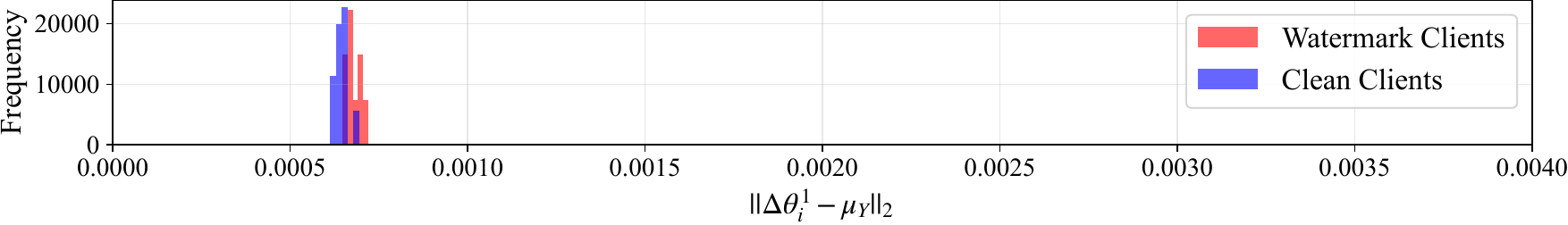}
    \caption{\Kuditipudi}
    \label{fig:tsne_p_allgen_dist}
    \end{subfigure}
    \caption{The distribution of the $\ell^2$-norm of displacement vector from each update to clean mean in the first round. All clean clients use synthetic data.}
    \label{fig:dist_diff-type}
\end{figure*}

For each update $\updates{i}{1}$ from clients in the first round, we measure the $\ell^2$-norm of the displacement from the clean mean, $\lVert \updates{i}{1} - \mu_Y \rVert_2$. Figure~\ref{fig:dist_diff-type} shows the distribution of these $\ell^2$-norms for \Kirchenbauer\ and \Kuditipudi\ in the first round when clean clients use synthetic data. From Figure~\ref{fig:tsne_k_allgen_dist}, we observe that, even with synthetic data for clean clients, \Kirchenbauer\ tends to produce larger $\ell^2$-norms for watermarking clients. Thus, these updates can still be filtered out in the first round. By contrast, \Kuditipudi\ induces little separation, and the distributions for clean and watermarking clients are very similar.

\section{Related Work}
\label{sec:relate-work}

\subsection{LLM Watermarks}

\citet{zhao2025sok} identify two primary families of LLM watermarks. We explore one representative watermark from each.

\paragraph{Green-Red Watermark.} Green-red watermark randomly partitions the vocabulary into ``green'' and ``red'' lists at each generation step and skews logits to favor green-listed tokens. Detection is performed via a statistical testing of the green-list token proportion in the generated text. \citet{kirchenbauer2023watermark} is one representative work in this family. It utilizes a hashing method to pseudo-randomly determine green list at each step, compared to approaches like \citet{zhao2023provable}, which predetermine a green list for all tokens. Subsequent research in this family focuses on addressing key limitations, including improving watermark robustness against adversarial attacks~\citep{liu2024semantic}, enhancing the watermark performance in low-entropy scenarios~\citep{lee2024wrote}, and increasing information capacity to carry multi-bit information~\citep{wang2024towards}.

\paragraph{Gumbel Watermark.} 
\citet{kuditipudi2023robust} is a representative watermark from the Gumbel watermark family. It employs a secret random number sequence to manipulate token sampling while preserving the original output distribution. The watermark is statistically distortion-free and does not rely on hashing. \citet{aaronson2023watermarking} proposes an alternative approach, which is hashing-based and computationally distortion-free provided that the tokens used for hashing do not repeat~\citep{zhao2025sok}. We select \citet{kuditipudi2023robust} over \citet{aaronson2023watermarking} because the latter's distortion-free guarantee depends on a precondition that may not hold in practice. Further analysis and experimental results for \citet{aaronson2023watermarking}'s approach are provided in Appendix~\ref{appendix:other}. As expected, experiments show that \citet{aaronson2023watermarking} is not robust to $\randeigen$.

\citet{sander2024watermarking} shows that hashing-based watermarks can exhibit {\em radioactive} properties in centralized setup. \citet{sander2024watermarking} also shows that continued fine-tuning on non-watermarked (clean) data can substantially reduce watermark detectability. We adapt their detection method and investigate if the findings still hold in FL setup. We additionally include an active server threat model and show that it can effectively remove the watermarked updates.

\subsection{Federated Learning}

\paragraph{FedSGD vs. FedAvg.} Federated learning (FL)~\citep{pmlr-v54-mcmahan17a} is a distributed machine learning framework where multiple clients collaboratively train a shared model under the coordination of a central server. Standard FL setting, FedSGD~\citep{pmlr-v54-mcmahan17a}, performs a single batch gradient over the entire local dataset per communication round. While straightforward, this approach typically requires a large number of communication rounds to converge, incurring high communication costs. FedAvg~\citep{pmlr-v54-mcmahan17a} significantly reduces communication overhead by running multiple local epochs on clients before synchronizing with the server. 

\paragraph{FL Optimization.} FedSGD and FedAvg rely on SGD for both local client updates and server aggregation, which can lead to slow convergence, particularly in scenarios involving non-IID data distributions~\citep{karimireddy2020scaffold} or environments with heavy-tailed random gradient noise distribution~\citep{zhang2020adaptive}. To address these challenges, many research explored the use of adaptive optimization methods to accelerate convergence. Several studies propose local adaptive optimization strategies, such as adaptively adjusting local learning rate~\citep{xie2019local, sun2023fedlalr}, maintaining local momentum buffers~\citep{yu2019linear}, or applying adaptive local interval~\citep{spiridonoff2020local, ji2020history}. FedOpt~\citep{reddi2020adaptive} introduces adaptive server-side optimizers (e.g., ADAGRAD, ADAM, or YOGI), which we adopt as our FL framework due to its significant improvement in convergence speed. Specifically, in some of our experiments, while FedAvg requires over 200 rounds to converge, FedOpt achieves convergence in less than 50 rounds.

\subsection{Byzantine-robust Aggregation}

FL systems that rely on a centralized server to simply average local client updates have been shown to be vulnerable to various adversarial attacks~\citep{bagdasaryan2020backdoor, bhagoji2019analyzing, nasr2019comprehensive, sun2021data}. Even a small number of malicious clients can stealthily distort the global model through carefully crafted updates~\citep{fang2020local}. Therefore, numerous Byzantine-robust FL protocols have been proposed to mitigate the impact of such malicious updates.

\paragraph{Weak Robust Aggregators.} Weak robust aggregators~\citep{NIPS2017_f4b9ec30, chen2020distributed, yin2018byzantine} compute measures of centrality per dimension and provide weak theoretical guarantees on the maximum bias. \citet{yin2018byzantine} propose aggregating local updates using coordinate-wise trimmed mean, while others leverage geometric median~\citep{chen2020distributed} or Euclidean distances between vectors~\citep{NIPS2017_f4b9ec30}). When applied to an $\wmratio$-corrupted set of $d$-dimensional vectors, none of these methods can achieve a total bias bound tighter than $O(\sqrt{\wmratio d})$~\citep{lai2016agnostic, lugosi2021robust}. 

\paragraph{Strong Robust Aggregators.} Strong robust aggregators compare magnitudes across all possible vector directions. They provides strong bias bounds independent of $d$ — a crucial security guarantee for high-dimensional ML models. Polynomial-time strong aggregators~\citep{diakonikolas2017being, hopkins2020robust, kothari2017outlier} identify outliers by examining their projections onto the dominant eigenvectors of the sample covariance matrix. 
While these polynomial-time methods offer strong theoretical guarantees, they often cannot achieve these bounds in practice due to computational limitations. Specifically, these methods require $O(\wmratio d^2)$ operations and $O(\ d^2)$ memory, rendering them impractical and vulnerable to attacks in high-dimensional settings~\citep{choudhary2024attacking,lee2025practical}. 

We employ $\randeigen$~\citep{lee2025practical} in our $\activeFL$, which is a state-of-art strong Byzantine robust aggregator that runs in quasi-linear running time, $O(Nd)$, and possesses provably near-optimal bias bounds. Unlike polynomial-time strong aggregators, $\randeigen$ does not require computing the maximum variance of clean vectors, $\|\Sigma_C\|_2$. Such design allows us to process an entire layer of the global model at once. In contrast, polynomial-time strong aggregators require splitting each layer into chunks of 1000 elements to meet practical memory constraints. Note that even with $\randeigen$, we must partition large layers (e.g. embedding layers of Pythia-70M) by factors of 8 or 16 to avoid memory overflow on an 80GB H100 GPU.

\subsection{Other Watermarking in FL}
While concurrent work also explores client-side FL watermarking scheme, WAFFLE ~\citep{yang2023watermarking} adopts a backdoor-based approach tailored for image classification tasks. In contrast, our work focuses on data-based watermarks for NLP tasks, addressing unique challenges in text-based FL environments.
%%%%%%%%%%%%%%%%%%%%%%%%%%%%%%%%%%%%%%%%%%%%%%%%%%%%%%%%%%%%

\newpage
\section*{NeurIPS Paper Checklist}
\begin{enumerate}

\item {\bf Claims}
    \item[] Question: Do the main claims made in the abstract and introduction accurately reflect the paper's contributions and scope?
    \item[] Answer: \answerYes{} 
    \item[] Justification: The abstract and Section~\ref{sec:intro} state our contributions, which are directly supported by the experimental results in Section~\ref{sec:eval}.

\item {\bf Limitations}
    \item[] Question: Does the paper discuss the limitations of the work performed by the authors?
    \item[] Answer: \answerNo{}
    \item[] Justification: Trade-offs and scope restrictions are discussed throughout the paper (e.g., Sections~\ref{sec:fed-prov-problem} and~\ref{sec:eval}). Due to space constraints, we do not include a dedicated limitation paragraph in the current draft. Additional limitations worth mentioning include experimental setups yet to be tested, such as the non-IID setting and larger LLMs. We will add a dedicated limitation paragraph before camera-ready.

\item {\bf Theory assumptions and proofs}
    \item[] Question: For each theoretical result, does the paper provide the full set of assumptions and a complete (and correct) proof?
    \item[] Answer: \answerNA{}
    \item[] Justification: The paper is primarily empirical. We introduce a formal  definition of FL watermark robustness (Definition~\ref{def:robustness}) and rely on existing bias-bound guarantees for strong robust aggregators from prior work~\cite{lee2025practical}, but we do not prove new theorems.

    \item {\bf Experimental result reproducibility}
    \item[] Question: Does the paper fully disclose all the information needed to reproduce the main experimental results of the paper to the extent that it affects the main claims and/or conclusions of the paper (regardless of whether the code and data are provided or not)?
    \item[] Answer: \answerYes{}
    \item[] Justification: All information required to reproduce the experimental results is provided. 

\item {\bf Open access to data and code}
    \item[] Question: Does the paper provide open access to the data and code, with sufficient instructions to faithfully reproduce the main experimental results, as described in supplemental material?
    \item[] Answer: \answerYes{}
    \item[] Justification: See Appendix~\ref{append:reproducibility}.

\item {\bf Experimental setting/details}
    \item[] Question: Does the paper specify all the training and test details (e.g., data splits, hyperparameters, how they were chosen, type of optimizer) necessary to understand the results?
    \item[] Answer: \answerYes{}
    \item[] Justification: All experimental details can be found in either Section~\ref{sec:eval} or the Appendices.

\item {\bf Experiment statistical significance}
    \item[] Question: Does the paper report error bars suitably and correctly defined or other appropriate information about the statistical significance of the experiments?
    \item[] Answer: \answerNo{}
    \item[] Justification: We report watermark detection $p$-values, but we do not report variability across multiple independent training runs. We acknowledge this as a limitation.

\item {\bf Experiments compute resources}
    \item[] Question: For each experiment, does the paper provide sufficient information on the computer resources (type of compute workers, memory, time of execution) needed to reproduce the experiments?
    \item[] Answer: \answerNo{}
    \item[] Justification: We report the GPU model and memory capacity used for the experiments (80GB H100 GPU), but do not provide complete per-experiment runtime or total compute estimates.
    
\item {\bf Code of ethics}
    \item[] Question: Does the research conducted in the paper conform, in every respect, with the NeurIPS Code of Ethics \url{https://neurips.cc/public/EthicsGuidelines}?
    \item[] Answer: \answerYes{}
    \item[] Justification: As stated in Appendix~\ref{append:ethics}, we conform, in every respect, with the NeurIPS Code of Ethics. 

\item {\bf Broader impacts}
    \item[] Question: Does the paper discuss both potential positive societal impacts and negative societal impacts of the work performed?
    \item[] Answer: \answerYes{}
    \item[] Justification: We discussed in Appendix~\ref{append:ethics}.
    
\item {\bf Safeguards}
    \item[] Question: Does the paper describe safeguards that have been put in place for responsible release of data or models that have a high risk for misuse (e.g., pre-trained language models, image generators, or scraped datasets)?
    \item[] Answer: \answerYes{}
    \item[] Justification: As stated in Appendix~A, we explicitly avoid releasing any tool meant to remove third-party watermarks; the released code will only reproduce our benign FL and evaluation pipeline, not a standalone watermark-evasion utility.

\item {\bf Licenses for existing assets}
    \item[] Question: Are the creators or original owners of assets (e.g., code, data, models), used in the paper, properly credited and are the license and terms of use explicitly mentioned and properly respected?
    \item[] Answer: \answerNo{}
    \item[] Justification: We cite the original sources for all datasets and models used, but do not explicitly document the version and license of every asset in the paper.

\item {\bf New assets}
    \item[] Question: Are new assets introduced in the paper well documented and is the documentation provided alongside the assets?
    \item[] Answer: \answerNA{}
    \item[] Justification: The paper does not release a new dataset/model.

\item {\bf Crowdsourcing and research with human subjects}
    \item[] Question: For crowdsourcing experiments and research with human subjects, does the paper include the full text of instructions given to participants and screenshots, if applicable, as well as details about compensation (if any)? 
    \item[] Answer: \answerNA{}
    \item[] Justification: This paper does not involve crowdsourcing or research with human subjects.

\item {\bf Institutional review board (IRB) approvals or equivalent for research with human subjects}
    \item[] Question: Does the paper describe potential risks incurred by study participants, whether such risks were disclosed to the subjects, and whether Institutional Review Board (IRB) approvals (or an equivalent approval/review based on the requirements of your country or institution) were obtained?
    \item[] Answer: \answerNA{}
    \item[] Justification: This paper does not involve human subjects; no IRB approval was required.

\item {\bf Declaration of LLM usage}
    \item[] Question: Does the paper describe the usage of LLMs if it is an important, original, or non-standard component of the core methods in this research? Note that if the LLM is used only for writing, editing, or formatting purposes and does \emph{not} impact the core methodology, scientific rigor, or originality of the research, declaration is not required.
    %this research? 
    \item[] Answer: \answerYes{}
    \item[] Justification: As stated in Appendix~\ref{append:LLM}, an LLM was used solely to polish grammar and clarity of author-drafted text and did not contribute to research ideation, experimental design, analysis, or the generation of novel content.

\end{enumerate}
\end{document}